\documentclass[12pt,preprint]{aastex}

\begin{document}
\title{Detectability of Occultation of Stars by Objects in the Kuiper
Belt and Oort Cloud} \author{T.~C.~Nihei\altaffilmark{1,2},
M.~J.~Lehner\altaffilmark{2} F.~B.~Bianco\altaffilmark{1,2},
S.-K.~King\altaffilmark{3}, J.~M.~Giammarco\altaffilmark{4},
C.~Alcock\altaffilmark{2} } \altaffiltext{1}{Department of Physics and
Astronomy, University of Pennsylvania, 209 South 33rd Street,
Philadelphia, PA 19104} \altaffiltext{2}{Harvard-Smithsonian Center
for Astrophysics, 60 Garden Street, Cambridge, MA 02138}
\altaffiltext{3}{Institute of Astronomy and Astrophysics, Academia
Sinica, P.O. Box 23-141, Taipei 10617, Taiwan}
\altaffiltext{4}{Department of Physics, Temple University, Barton
Hall, Philadelphia, PA 19122}

\begin{abstract}
The serendipitous detection of stellar occultations by Outer Solar
System objects is a powerful method for ascertaining the small end
($r\lesssim 15$~km) of the size distribution of Kuiper Belt
Objects and may potentially allow the exploration of objects as far
out as the Oort Cloud.  The design and implementation of an
occultation survey is aided by a detailed understanding of how
diffraction and observational parameters affect the detection of
occultation events.  In this study, stellar occultations are
simulated, accounting for diffraction effects, finite source sizes,
finite bandwidths, stellar spectra, sampling, and
signal-to-noise. Finally, the possibility of detecting small Outer
Solar System objects from the Kuiper Belt all the way out to the Oort
Cloud is explored for three photometric systems: a proposed space
telescope, Whipple \citep{Kaplan2003}, the Taiwanese-American
Occultation Survey \citep{Lehner2006}, and the Multi Mirror Telescope
\citep{Bianco2007}.
\end{abstract}


\section{Introduction}
Stellar occultation, the dimming of a background star by a foreground
object passing through the line of sight, is used in a variety of
scientific studies to probe the properties of foreground objects.
With stellar occultations, it has been possible to discover and study
planetary rings \citep[e.g.][references therein]{Bosh2002} and the
atmospheres of planets and their satellites
\citep[e.g.][]{Elliot2003a, Elliot2003b, Gulbis2006, Pasachoff2005,
Sicardy2006}.

Studies suggest that by searching for serendipitous occultations
in monitored stars it may be possible to detect Outer Solar Systems
objects in the Kuiper Belt \citep{Dyson1992, Axelrod1992} and objects
as far out as the inner edge of the Oort Cloud
\citep{Bailey1976}. This provides a novel way to ascertain the
size distribution of small object populations in the Kuiper Belt and
the Inner Oort Cloud.

The standard planet formation scenario begins with a disk of small
planetesimals (with radii $r < 0.5$~km) surrounding a protostar.
These planetesimals collide with one another and merge to become
larger objects.  When sufficiently massive, gravitational focusing
leads to runaway accretion.  Larger objects dominate the accretion
process and eventually go on to become planets.  Remaining
planetesimals, by a variety of mechanisms, are cleared from the planet
region of the disk.  Some objects are ejected to distant orbits by
perturbations from the giant planets.  At the outer edge of the
protostellar disk, a slow rate of collision fails to produce large
enough objects and planet formation fails to occur. Far enough removed
from the massive planets, these smaller planetesimals are invulnerable
to perturbations and remain in the disk.

In our own Solar System, the Kuiper Belt is a remnant of the outer
protostellar disk which failed to form planets.  The population of
planetesimals in this region have most likely been perturbed by Neptune
and perhaps other massive bodies, therefore the Kuiper Belt's size
distribution, spatial distribution, and mass are important keys to
understanding the evolution of planetary disks.

Models and observations suggest the differential size distribution of
the Kuiper Belt follows a broken power law $N \propto r^{-q}$, where
$q$ takes on different values in the lower and upper size regions.
The location of the break $r_\mathrm{b}$ in this power law depends on
the initial mass, size spectrum, and bulk properties of Kuiper Belt
Objects (KBOs), and Neptune's orbital evolution among other parameters
\citep{KandB2004, Kenyon2002, PandS2005, Stern1996}.  Observations
have constrained the size spectrum for large objects ($r \gtrsim
100$~km) to an index $q \sim 4$ \citep{Bernstein2004, Trujillo2001}.
\cite{Bernstein2004} has found evidence for a break in the size
distribution near $r_\mathrm{b} \sim 30$--50~km by a faint object
survey mounted on the Advanced Camera for Surveys (ACS) on the Hubble
Space Telescope (HST).  Previous models predicted the location of the
break radius at $0.05~\mathrm{km}<r_\mathrm{b}<5~\mathrm{km}$
\citep{Stern1996, KandL1999, Kenyon2002}. \cite{KandB2004} and
\cite{PandS2005} have since modeled the Kuiper Belt size distribution
revisiting assumptions made about the relative gravitational and
tensile energies of KBOs.  \cite{PandS2005} concluded that bulk
strength plays little role in the fragmentation of small objects and
find reasonable agreement with the suggested break location evidenced
by \cite{Bernstein2004}.  \cite{KandB2004} estimated the break radius
at $0.5~\mathrm{km}<r_\mathrm{b}<15~\mathrm{km}$.  Additionally they
suggested that a dip should be present at $5<r<35$~km due to the
removal of small objects due to collisional erosion.  These two models
differ in the small size range where direct observations are difficult
due to the dim surface brightness of KBOs smaller than $r = 15$~km.

Surveys monitoring background stars for occultations due
to KBOs have the capability to determine the small end of the KBO size
distribution \citep{RandM2000, CandF2003}.  Various groups are now
attempting to implement such surveys \citep[e.g.][]{Roques2003,
Bickerton2006, Lehner2006, Chang2006}.  \cite{Roques2003} have
implemented a survey at the Pic du Midi Observatory with
sampling at 20~Hz frequency. They reported a candidate event
at $3\sigma$ which may be ascribed to an occultation by a $r\sim
0.15$~km KBO.  Three candidate occultation events were also reported
by a later survey conducted with a frame transfer camera mounted on
the 4.2~m William Herschel Telescope at La Palma (observing at
46~Hz) \citep{Roques2006}.  However, the survey team ruled out the
detection of any KBOs in the $30$--50~AU range.  \cite{Bickerton2006}
have also followed suit by implementing a high-speed CCD camera design
with 40~Hz cadence and mounting a search with the 72-inch Plaskett
telescope in Victoria, BC. Additionally, \cite{King2002} and
\cite{Lehner2006} have described a dedicated occultation survey known
as The Taiwanese-American Occultation Survey (TAOS) which uses three
wide-field robotic telescopes to monitor as many as 2,000 stars for
chance occultations by KBOs.

Recently, candidate occultation events at millisecond timescales were
observed in X-ray lightcurves of Scorpius X-1 \citep{Chang2006}.
These events were claimed to be compatible with KBOs in the size range
$5~\mathrm{m}< r <100~\mathrm{m}$.  The reported event rate is much
higher than expected from models of KBO formation.  \cite{Jones2006}
have argued that the candidate events observed by \cite{Chang2006} are
attributed to dead time response due to charged particle
scintillation.  Subsequently, \cite{Chang2007} adjusted their number
of candidate occultation events to account for this effect.  To date,
occultation surveys have been unable to provide definite constraints
on the small size distribution of objects in the Kuiper Belt.

Beyond the Kuiper Belt, there remain many questions that could be addressed by
stellar occultation surveys.  The structure of the Outer Solar System
is thought to extend from the Kuiper Belt out to distances as large as
$\sim 100,000$~AU.  The nature of this vast region holds
important clues to the complex evolution of our Solar System.

The Oort Cloud is a population of planetesimals that is thought to
have been scattered from the planetary disk roughly $\sim 4.5$~Gyr
ago, and is predicted to have a wide range of orbits with semi-major
axes spanning distances of $1,000~\mathrm{AU} < a <
100,000~\mathrm{AU}$.  Its existence was proposed as a probable
reservoir of long-period comets \citep{Oort1950}.

Models suggest that these objects formed in the region of the giant
planets ($\sim 4-40$~AU) and were perturbed by the giant planets to
large orbits with relatively unchanged perihelia. Interactions with
nearby passing stars, giant molecular clouds, and other material in
the solar neighborhood would have increased their perihelia to larger
distances, increased the inclinations, and placed them in large orbits
to form the Oort Cloud \citep{Dones2004}.  Dynamical simulations
combined with the flux of long-period comets into the planetary region
of the Solar System lead to estimates of the number of comets in the
Oort Cloud to be $\sim 10^{11}-10^{12}$.  However, because the Oort
Cloud is at such large distances and its comets are expected to be the
size of observed long-period comet nuclei ($r\sim 3$~km), direct
observations of the Oort Cloud are unlikely.

The discovery of Sedna (2003 VB12), a $r = 830$~km object on an
eccentric orbit at heliocentric semi-major axis of $a = 490$~AU, came
as a surprise \citep{Brown2004}.  An object of substantial size at
such a distant orbit was unexpected.  Sedna's origins remain unclear,
but its discoverers \citep{Brown2004} have suggested that Sedna may be
part of the Oort Cloud.  This implies that Sedna amassed size in the
planetary region and was perturbed to its current orbit
\citep{MandL2004, Matese2005}.  \cite{Stern2005} has pointed out that
Sedna's eccentric and inclined orbit does not preclude its formation
\emph{in situ}. If this is true, it is possible that Sedna is part of
a population of objects that lie in a proposed annular region beyond
the Kuiper Belt \citep{Brasser2006} which is
referred to in this paper as the Extended Disk ($a \sim
50$--1,000~AU).

Using stellar occultations to search for Extended Disk objects is also
a possibility.  To date, no known occultation surveys are dedicated to
this purpose.  However, \cite{Roques2006} showed that this may be a
real possibility as they identified two candidate occultation
events by 300~m radius objects beyond 100~AU via this
method.

The potential for occultation surveys to detect objects in the Kuiper
Belt is well discussed in the literature.  However, the capability of
a given occultation survey to detect objects out to the Oort Cloud
still requires careful consideration as it depends on several factors.
In particular, \cite{RandM2000} have shown that star light diffraction
must be taken into account for surveys looking to detect stellar
occultations by small objects in the Kuiper Belt.  In this paper, the
effects of stellar types, finite bandwidth and sampling on a survey's
ability to detect occultation events in the Outer Solar System are
studied.  Occultation events for three photometric systems are
simulated in order to guide the selection of observational parameters
for occultation surveys. In \S~\ref{sec:diffraction} a brief
introduction to stellar occultation and diffraction effects are given.
The effects of finite bandwidth, stellar spectra, and finite source
sizes are considered later in \S~\ref{sec:finite}.  The effects of
sampling is discussed in \S~\ref{sec:velocity}.  Noise for three
photometric systems as mentioned above is included in
\S~\ref{sec:noise}. Finally, the detectability of an occultation event
for photometric systems and the conclusions of this study are
discussed in \S~\ref{sec:detect}, \S~\ref{sec:threshold}, and
\S~\ref{sec:discuss}.

\section{Stellar Occultations and Diffraction}
\label{sec:diffraction}
The diffraction pattern created by the stellar occultation of a
distant star by a foreground spherical object is described using
Lommel functions,
\begin{displaymath}
U_{n}(\mu, \nu) = \sum\limits^{\infty}_{k=0}\left( -1\right)^k
\left(\frac{\mu}{\nu}\right)^{n+2k} J_{n+2k}(\pi \mu \nu),
\end{displaymath}  
where $J_n$ is a Bessel function of order $n$.  For the case of an
occultation by an object of radius $r$ at a distance
$a$, the measured intensity of a star at
wavelength $\lambda$ is described \citep[and references
therein]{Roques1987} by
\begin{eqnarray}
I_{\rho}(\eta)\!=\!\left\{\!\!
  \begin{array}{ll}
  U_{0}^{2}(\rho,\eta) + U_{1}^{2}(\rho, \eta)\vspace{0.05in} & 
  \eta \leq \rho \vspace{0.05in}\\ 
  1 + U_{1}^{2}(\rho,\eta) + U_{2}^{2}(\rho, \eta) & \eta \geq \rho
  \\ 
  \:- 2U_{1}(\rho,\eta)\sin\frac{\pi}{2}(\rho^{2} + \eta^{2}) & \\ 
  \:+ 2U_{2}(\rho,\eta)\cos\frac{\pi}{2}(\rho^{2} + \eta^{2}) &
  \end{array} 
  \right.,
\label{eq:occult_intensity}
\end{eqnarray}
where $\rho = r/F$ and $\eta= x/F$ are the radius and distance from
the line of sight in units of the \emph{Fresnel scale}
$F=\sqrt{\lambda a/2}$. The subscript on $I_\rho$ indicates that the
occultation pattern depends solely on the dimensionless parameter
$\rho$.

\begin{figure}
  \epsscale{1.0} \plotone{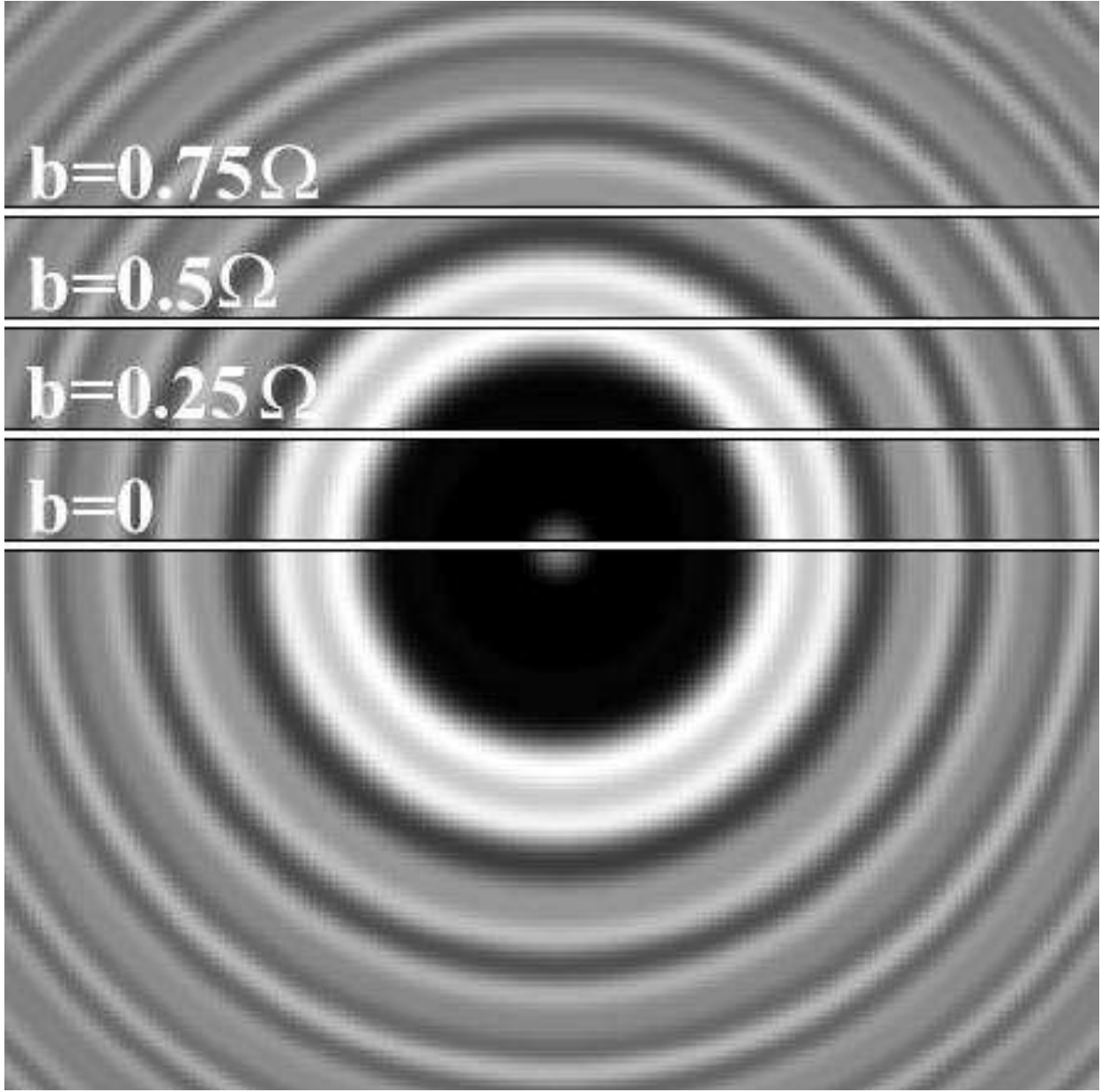}
  \caption{The projected occultation pattern for $\rho = 1$.  The
    scale of the pattern is 5 Fresnel units on a side.  Four
    trajectories crossing the occultation pattern with impact
    parameters $b$ are shown with corresponding intensity curves in
    Figure~\protect\ref{fig:imp}.  Values for the impact parameters
    are expressed in terms of the occultation width $\Omega$ shown in
    Figure~\protect\ref{fig:imp}. }
  \label{fig:diffproj}
\end{figure}

\begin{figure*}
 \epsscale{1.0} \plotone{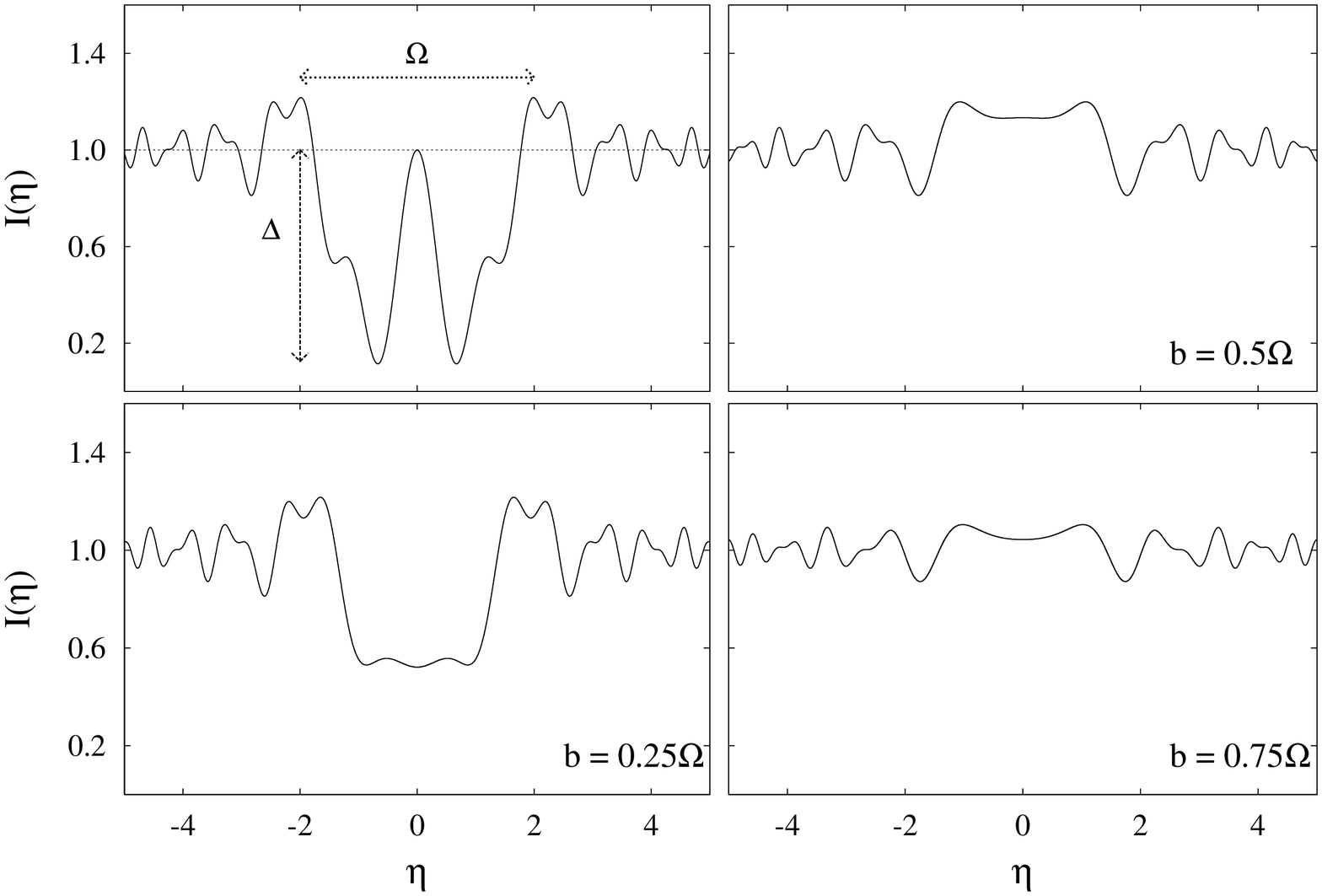}
  \caption{Intensity curves for an occultation pattern for $\rho=1$
    traversed at various impact parameters $b$.  Impact parameter
    values are labeled at the bottom of each plot and correspond to
    their labeled counterparts in Figure~\protect\ref{fig:diffproj}.}
\label{fig:imp}
\end{figure*}

Figure~\ref{fig:diffproj} shows a projected occultation pattern
computed from the above expression for $\rho=1.0$.  Also shown are
four trajectories through the occultation pattern at four different
values of the impact parameter $b$, given in terms of the diameter of
the first Airy ring (this is an approximate definition as will be
discussed further), which we will call $\Omega$. The corresponding
intensity profile curves for each of these trajectories are shown in
Figure~\ref{fig:imp}.  An object crossing a source with a finite
impact parameter leads to several possible intensity curves for a
given occultation pattern. From the plots shown in
Figure~\ref{fig:imp}, it can be seen that for impact parameters $b \ge
0.5\Omega$, the occultation event depth is quite small, and such
events will be very difficult to detect.  We thus define an
occultation \emph{event} as an object crossing the line of sight to a
star with an impact parameter $b \le 0.5\Omega$, and we define the
\emph{event width} as the first Airy ring diameter $\Omega$. Finite
impact parameters complicate the characterization of an occultation
event because of the multitude of possible intensity curves that may
arise from one occultation pattern. In order to simplify the
discussion, only the case of $b=0$ is considered throughout the
remainder of this paper. This assumption does not significantly alter
the conclusions.

Note that at center of the diffraction pattern in
Figure~\ref{fig:diffproj}, $I_{\rho}(0)=1$; this is the \emph{Poisson
spot} and is a consequence of diffraction present for circular objects
passing before a point source.

The parameter $\rho$ defines the strength of diffraction effects, in
particular the width and depth of the occultation. Assuming a point
source background star, the occultation width $\Omega$ (see
Figure~\ref{fig:imp}) takes on two values in the limiting cases of
large and small $\rho$.  For $\rho \ll 1$, the occultation pattern is
completely dominated by diffraction and the diameter of the Airy ring
is given by $2\sqrt{3}$ in dimensionless Fresnel scale units.  In the
case where $\rho\gg 1$, the diffraction effects become negligible and
the width approaches the limiting case of $2\rho$. An empirical
approximation over the entire region of interest is given by

\begin{equation}\label{eq:width_fsu}
\Omega \approx 2\left[\left(\sqrt{3}\right)^{3 \over 2}
+ \rho^{3 \over 2}\right]^{{2 \over 3}}.
\end{equation}
In physical units, the width $W$ is
\begin{equation}
W \approx 2 \left[\left(\sqrt{3}F\right)^{3 \over 2}
+ r^{3 \over 2}\right]^{{2 \over 3}}.
\end{equation}\label{eq:width} 

\begin{figure}
  \epsscale{1.0} \plotone{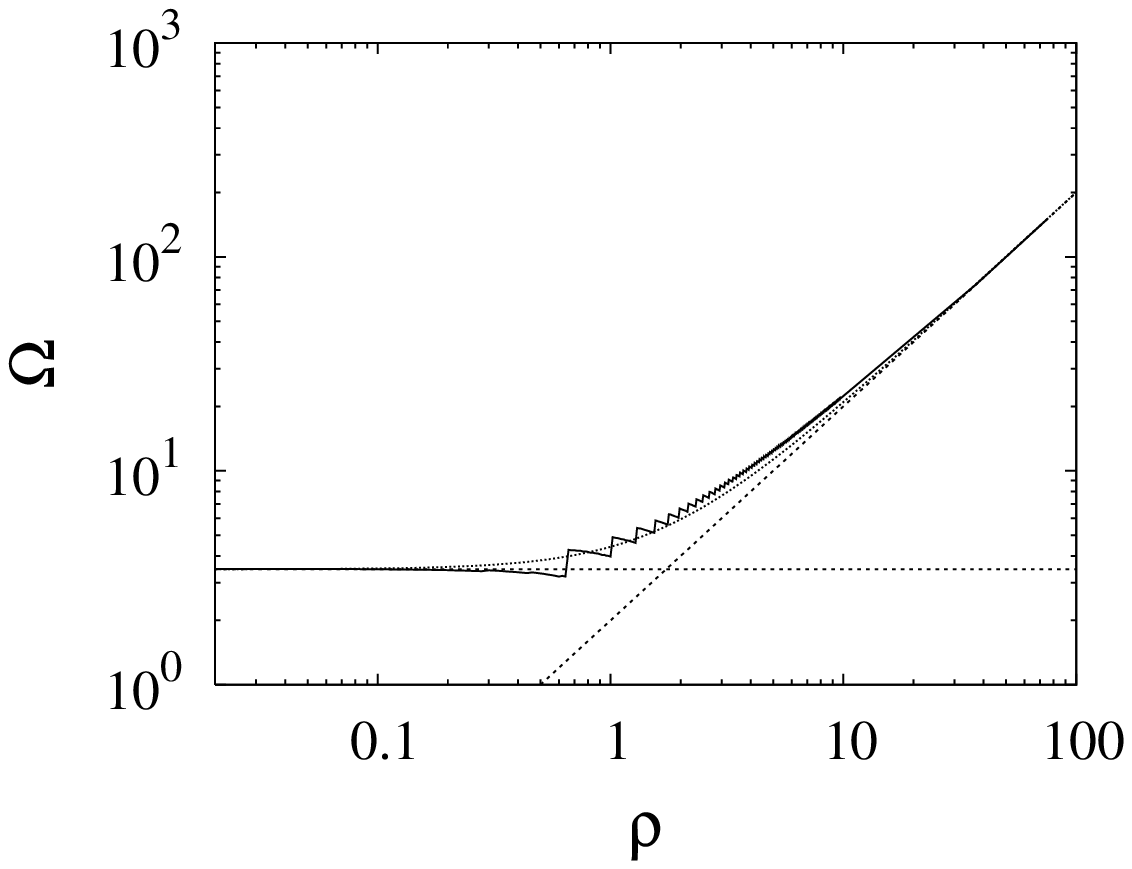}
  \caption{Occultation width $\Omega$ (see
  Figure~\protect\ref{fig:imp}) versus the object radius $\rho$.  The
  solid line shows the measured width, the dotted line shows the
  approximation given in Equation~\protect\ref{eq:width_fsu}, and the
  dashed lines indicate the asymptotic behavior for $\Omega$ in the
  two limiting cases for $\rho$.}\label{fig:eventwidth_fsu}
\end{figure}

\begin{figure}
  \epsscale{1.0} \plotone{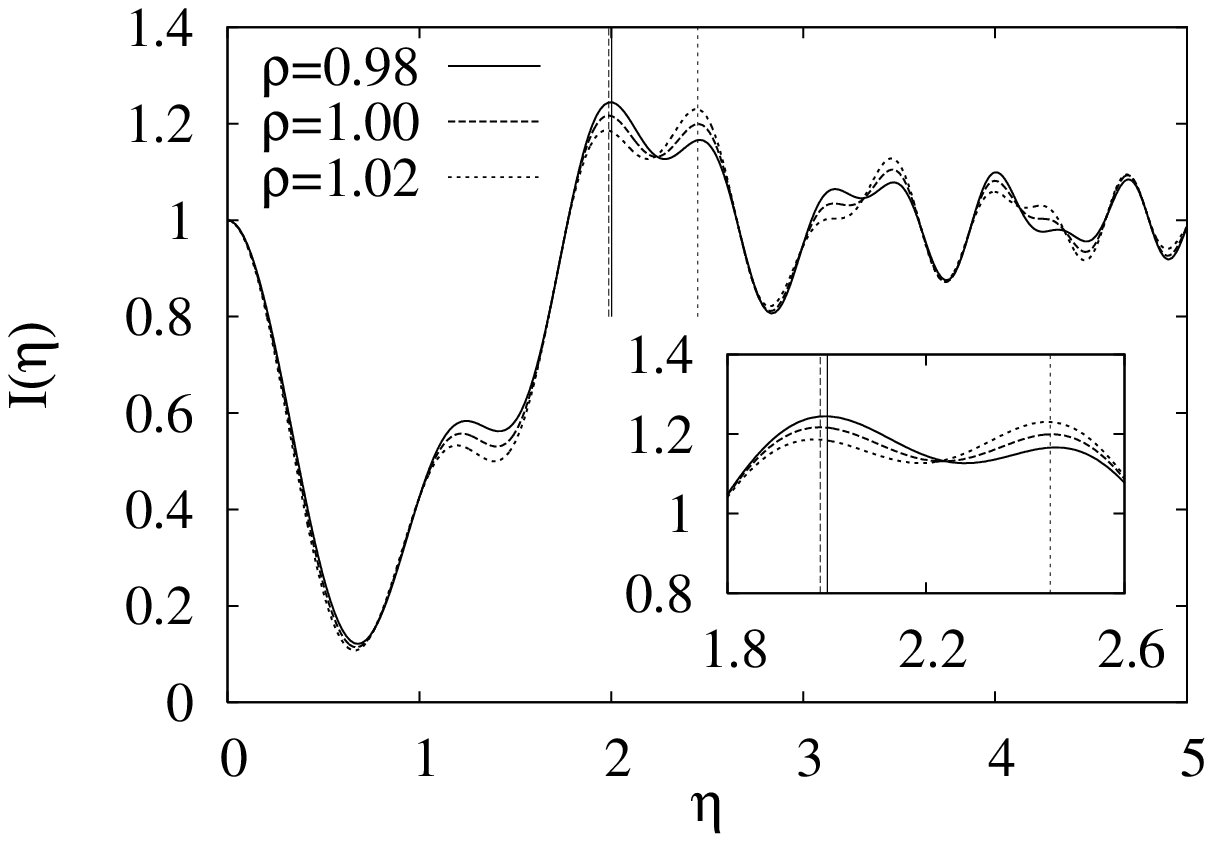}
  \caption{Occultation patterns for $\rho =0.98$, 1.00, and
  1.02. Vertical lines indicate the measured widths for each pattern.
  Increases in $\rho$ shift the absolute maximum peak by a significant
  amount.}
\label{fig:jump}
\end{figure}

Figure~\ref{fig:eventwidth_fsu} shows the measured width (solid line)
against the approximation in Equation~\ref{eq:width_fsu} (dashed
line).  The two limiting expressions for large and small $\rho$ are
indicated by dotted-lines.  The ``jumps'' in occultation width
$\Omega$ for $0.3 \lesssim \rho \lesssim 9$ are the result of the
gradual shifting of diffraction fringes near the shadow edge.
Figure~\ref{fig:jump} depicts three lightcurves near $\rho \sim 1$
($\rho = 0.98$, 1.00, and 1.02).  Determining the maximum peak is
complicated by diffraction fringes that span the shadow edge.  The
jumps in the event width are due to one of two local maxima becoming
larger than the other.  The measured width for each of the curves are
depicted in Figure~\ref{fig:jump} by vertical lines drawn to match the
appropriate curve.

For a source of finite angular radius $\theta_{\ast}$, the projected radius 
in the plane of the occulting object is $r_{\ast}= a\theta_{\ast}$.  
This finite radius extends the occultation width by the projected
diameter of the star such that 
\begin{equation}\label{eq:width_fsu_finite} 
\Omega^\ast \approx 2\left[\left(\sqrt{3}\right)^{3 \over 2}
+ \rho^{3 \over 2}\right]^{{2 \over 3}} + 2 \rho_{\ast},
\end{equation}
where $\rho_{\ast} = r_{\ast}/F$, and the asterisk in the superscript
indicates that a finite source disk is accounted for in the
occultation width.  The width in physical units is thus
\begin{equation}\label{eq:width_finite} 
W^\ast \approx 2 \left[\left(\sqrt{3}F\right)^{3 \over 2}
+ r^{3 \over 2}\right]^{{2 \over 3}} + 2 r_{\ast}.
\end{equation}

\begin{figure}
  \epsscale{1.0} \plotone{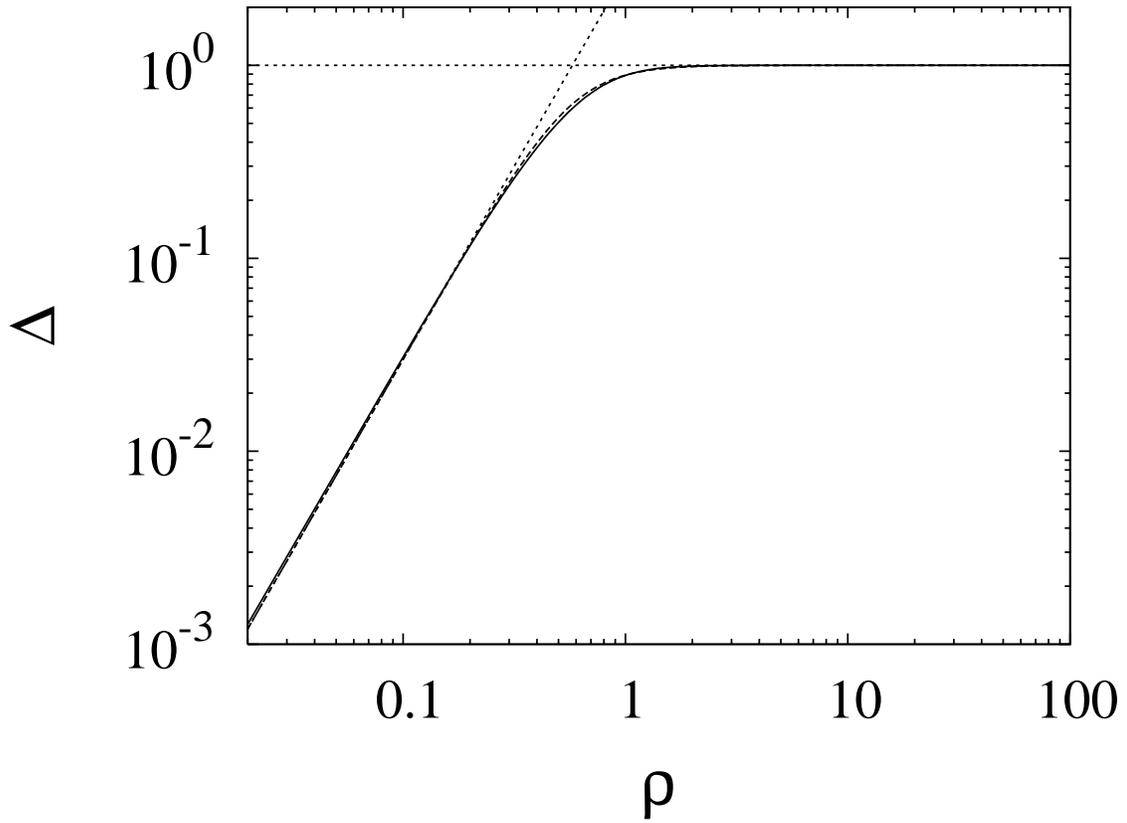}
  \caption{Occultation event depth $\Delta$ (see
  Figure~\protect\ref{fig:imp}), as function of $\rho$.  The solid
  line shows the measured depth, the dashed line shows the empirical
  approximation in Equation~\protect\ref{eq:depth}, and the dotted
  lines show the two limiting cases.}
  \label{fig:eventdepth_fsu}
\end{figure}

The occultation depth $\Delta$ is defined as the magnitude of the
maximum downward deviation of an occultation pattern, as drawn in
Figure~\ref{fig:imp}.  Figure~\ref{fig:eventdepth_fsu} is a plot of
the depth $\Delta$ as a function of $\rho$.  There are two limiting
cases of the occultation depth as a function of $\rho$. An empirical
fit to $\Delta(\rho)$ for $\rho \ll 1$ shows that the depth roughly
follows a power law $\Delta \approx 3\rho^{2}$.  In the limit where
$\rho \gg 1$, the object disk completely extinguishes the background
source, hence the depth is constant at $\Delta=1$.  By an empirical
fit to the measured depth, we arrive at the expression
\begin{equation}\label{eq:depth}
\Delta = \left[1 + \left(3 \rho^{2}\right)^{-{3 \over 2}}\right]^{-{2
  \over 3}}
\end{equation}

\begin{figure*}
 \epsscale{1.0} \plotone{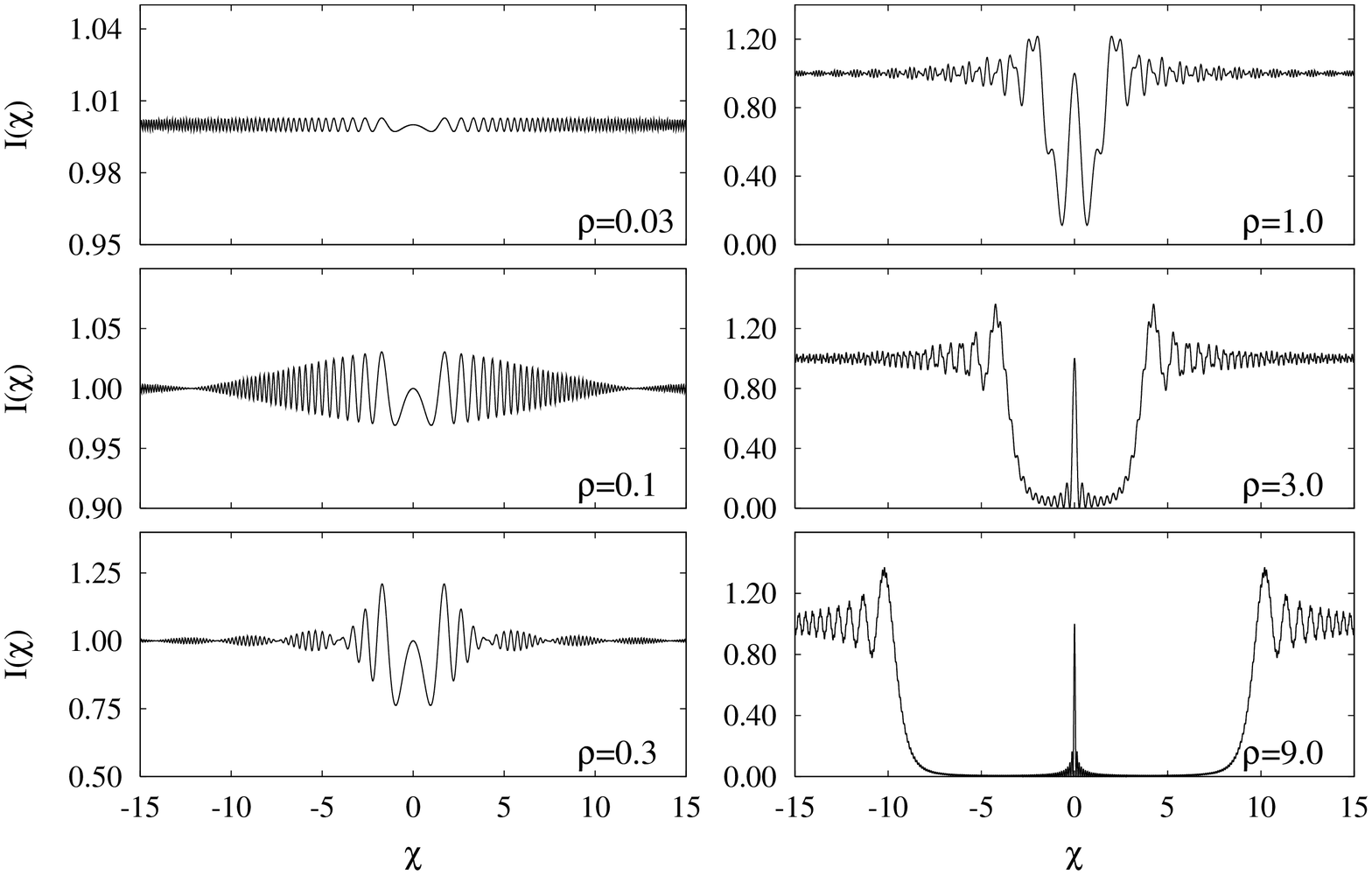}
  \caption{Occultation intensity profiles measured across the center
    of the pattern for several values of $\rho$.}
  \label{fig:fsu_evolve}
\end{figure*}

Three regimes are apparent in
Figures~\ref{fig:eventwidth_fsu}~\&~\ref{fig:eventdepth_fsu}: (1) the
\emph{far-field} or \emph{Fraunhofer} regime for $\rho\ll 1$, (2) the
\emph{near-field} or \emph{Fresnel} regime for $\rho\sim 1$, and (3)
the \emph{geometric} regime for $\rho\gg 1$.  In the Fraunhofer regime
the occultation width $\Omega = 2\sqrt{3}$ is independent of $\rho$,
but the depth varies as $\Delta \approx 3\rho^{2}$. Example
diffraction patterns in this regime are shown in
Figure~\ref{fig:fsu_evolve} for the cases $\rho = 0.03$, $0.01$, and
$0.3$. It is clearly evident in this figure that as $\rho$ decreases,
the width remains constant and the depth decreases.  In the Fresnel
regime the depth and width of the occultation event are both dependent
on $\rho$, hence as $\rho$ increases both parameters will increase as
seen in Figure~\ref{fig:fsu_evolve} where $\rho=1.0$, and $3.0$.
Finally, geometric patterns have a width which approaches the
geometric shadow $\Omega\to 2\rho$ and complete extinction of the
background source is apparent such that $\Delta = 1$.
Figure~\ref{fig:fsu_evolve} shows an occultation event approaching the
geometric regime in the panel for $\rho=9.0$.

\section{Spectral Type and Finite Size of Source Star}
\label{sec:finite}
Occultation patterns described in the previous section depend upon the
observation wavelength $\lambda$. In reality, a survey will monitor
stars through a finite bandwidth and this can affect the diffraction
features observed in an occultation pattern.  If the bandwidth is
narrow, a monochromatic pattern is a reasonable approximation for an
observed occultation.  For broader bandwidths like that of TAOS, we
describe the intensity pattern as
\begin{equation}
    I_{r,a}(x) = \int\limits^{\infty}_{-\infty}
    s(\lambda)f(\lambda)I_{r,a}(x,\lambda)
    d\lambda, 
\end{equation}
where 
\begin{eqnarray}
I_{r,a}(x, \lambda)& = & I_{\rho}(\eta) \nonumber\\
                   & = & I_{r/F(a,\lambda)}(x/F(a,\lambda)) \nonumber,
\end{eqnarray}
and the wavelength-dependent filter transmission and stellar spectrum
are represented by $f(\lambda)$ and $s(\lambda)$ respectively.

To determine broad bandwidth effects, $I_{r,a}(x,\lambda)$ for several
$r$ and $a$ were integrated against a flat spectrum with a series of
tophat filters.  Stellar spectra were accounted for using the UVILIB
library, a compiled stellar flux library spanning a total wavelength
range of 1,150--10,620~\AA\ \citep{Pickles1998}, was incorporated into
occultation curve simulations.  We found that for point sources
broader bandwidths will smooth and dampen diffraction fringes, while
broadening of the occultation width is minimally seen. Additionally,
stellar spectra appear to have little effect on the intensity profile
of a diffraction pattern for a background point source.

Throughout the remainder of this paper, the Fresnel scale of an event
with a broadband filter is redefined as
\begin{displaymath}
F = \sqrt{\bar\lambda a \over 2},
\end{displaymath}
where $\bar\lambda$ is the median wavelength of the filter of interest.

While the stellar spectra themselves have a minimal effect, finite
source sizes will significantly affect the occultation intensity
patterns, specifically when $\rho_{\ast} \gtrsim 1$. The source size
of a given star is determined by a combination of its stellar class
and apparent brightness.  To address finite source size effects,
stellar radii from tabulated values \citep{Cox2000, Lang1992} were
incorporated into simulations of occultation events.  (Limb darkening
effects also were considered using a solar limb darkening model
\citep{Cox2000}, however the effects were found to be $\lesssim 1$\%,
hence they are ignored throughout the rest of this discussion.)

The intensity profile from a finite source disk of projected radius
$r_{\ast}$ is expressed as an integral over the projected source disk
\begin{eqnarray}\label{eq:finsrc1}                              
I_{r,a}^{\ast}(x)& = &\frac{1}{\pi r^{2}_{\ast}} 
\int\limits^{2\pi}_{0}\int\limits_{0}^{r_{\ast}}
\left(r^{\prime}_{\ast}dr^{\prime}_{\ast}d\phi\right.
\nonumber\\        
& \times & 
\left.I_{r,a}(x^{2} + r^{\prime2}_{\ast} + 
2xr^{\prime}_{\ast}\cos\phi)\right).\nonumber
\end{eqnarray}                                
Here $r^{\prime}_{\ast}$ and $\phi$ are the distance from the center of the
source disk and the polar angle.  The asterisk in the superscript
$I^\ast$ denotes that the occultation pattern was calculated
accounting for the finite source size.

\begin{figure}
  \epsscale{1.0} \plotone{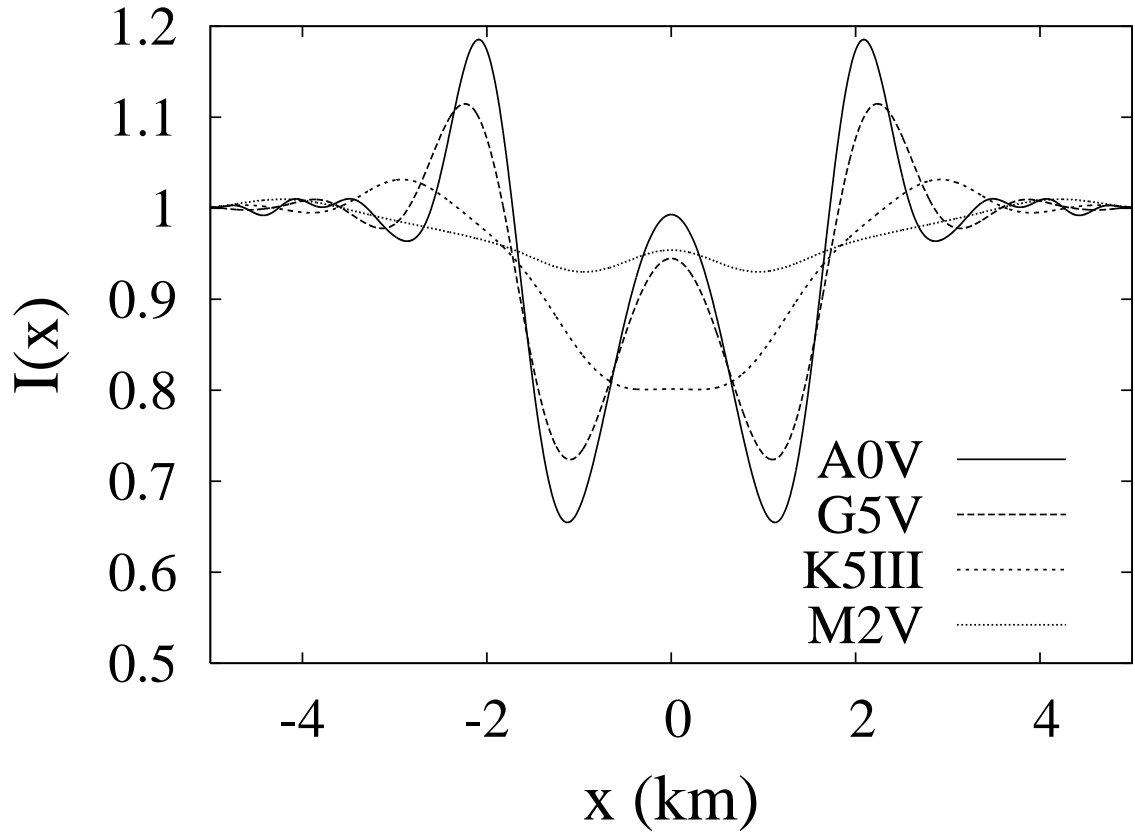}
  \caption{Diffraction profiles for a KBO for different $V=12$
    background stars: A0V, G5V, K5III, and M2V.  The occulting object
    for this set of curves has a radius of 0.5~km and is at a distance
    of 40~AU.  Projected stellar radii at 40~AU are 0.2, 0.5, 1.2, and
    2.6~km respectively.}
    \label{fig:finite_1km_40au}
\end{figure}

\begin{figure}
  \epsscale{1.0}
  \plotone{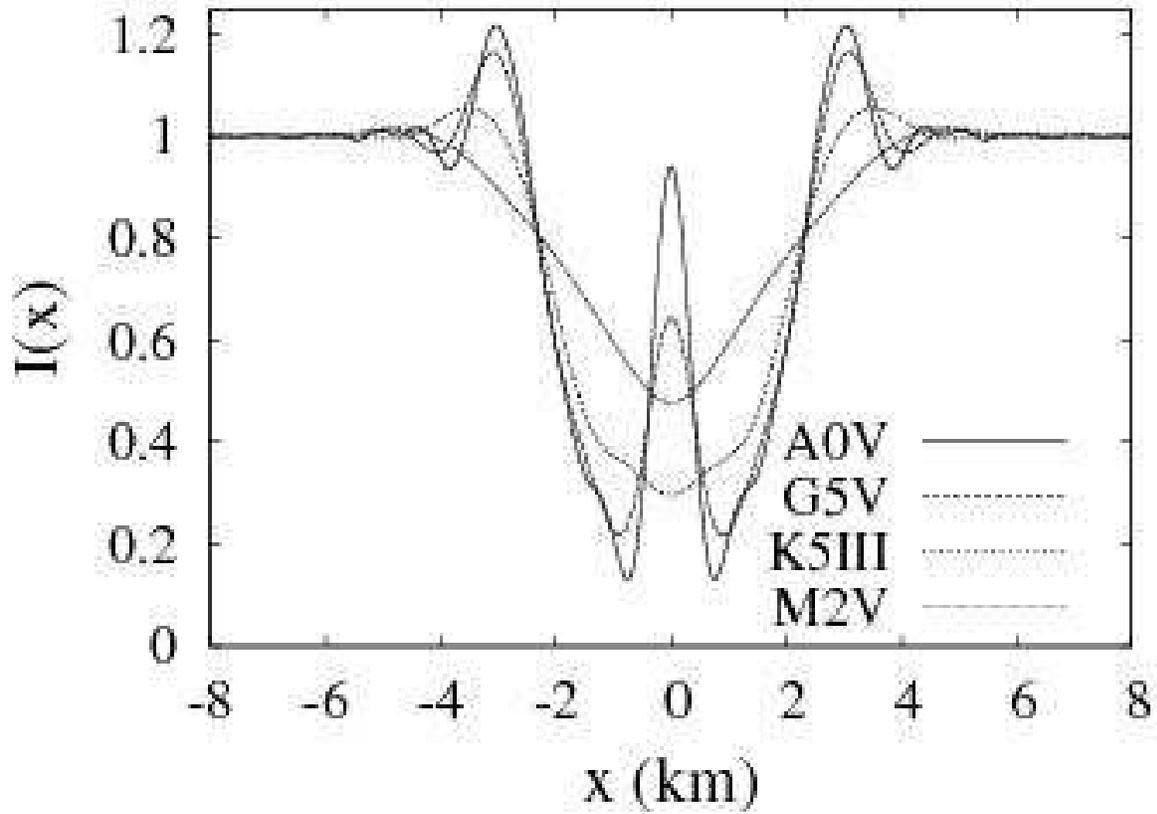} 
  \caption{Diffraction profiles for a KBO for different $V=12$
    background stars: A0V, G5V, K5III, and M2V.  The occulting object
    for this set of curves has a radius of 1.5~km and is at a distance
    40~AU.  Projected stellar radii at 40~AU are 0.2, 0.5, 1.2, and
    2.6~km respectively.}
  \label{fig:finite_3km_40au}
\end{figure}

Occultation profiles for four stellar types each with magnitude $V=12$
for $r= 0.5$ and 1.5~km objects are plotted in
Figures~\ref{fig:finite_1km_40au}~\&~\ref{fig:finite_3km_40au}.  Both
illustrate that the finite source size of a stellar disk can
significantly broaden the occultation width and smooth the diffraction
fringes.  In cases where the source size is relatively large such as
an M2V $V=12$ star, the smoothing is so significant that diffraction
fringes with variations as large as $\sim$35\% are reduced to
variations of $\sim$5\%.

\begin{figure}
  \epsscale{1.0} \plotone{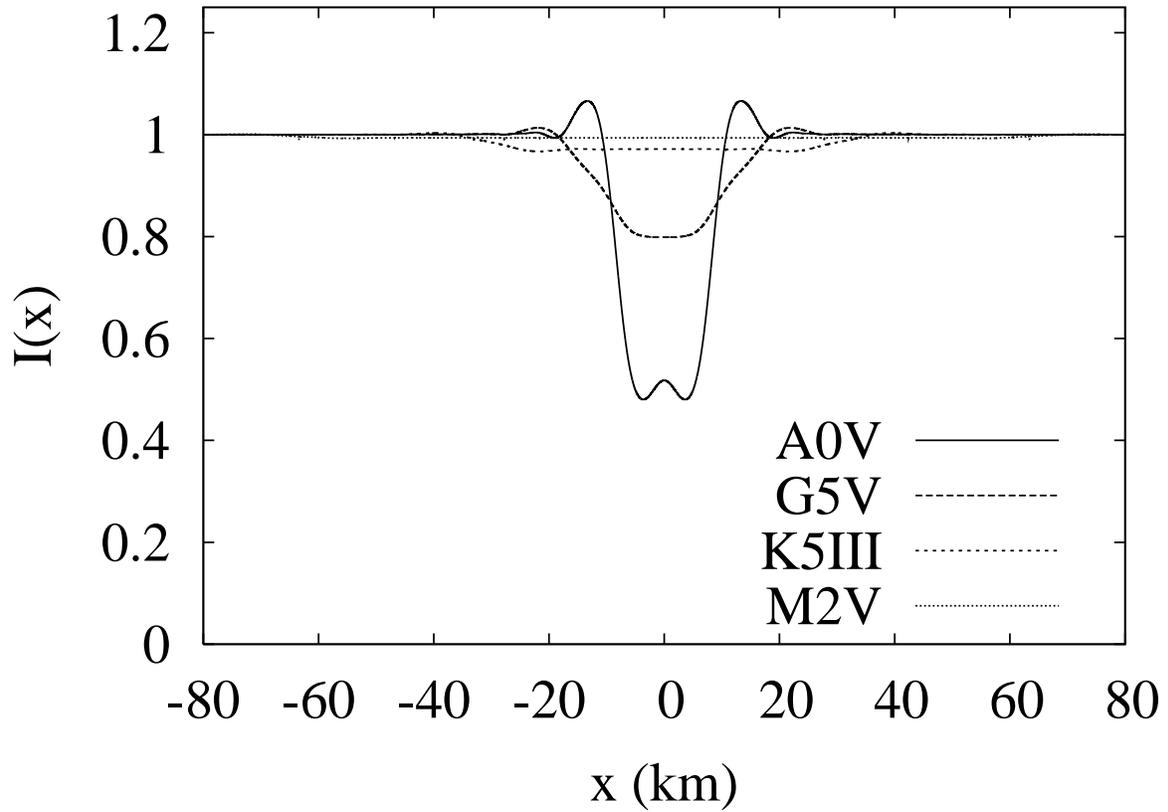}
  \caption{Diffraction profiles for an Extended Disk object for
    different $V=12$ background stars: A0V, G5V, K5III, and M2V.  The
    occulting object for this set of curves has a radius of 5~km and
    is at a distance of 1,000~AU.  Projected stellar radii at 1,000~AU
    are 4.3, 12.9, 30.6, and 64.1~km respectively.}
  \label{fig:finite_1km_1000au}
\end{figure}

\begin{figure}
  \epsscale{1.0} \plotone{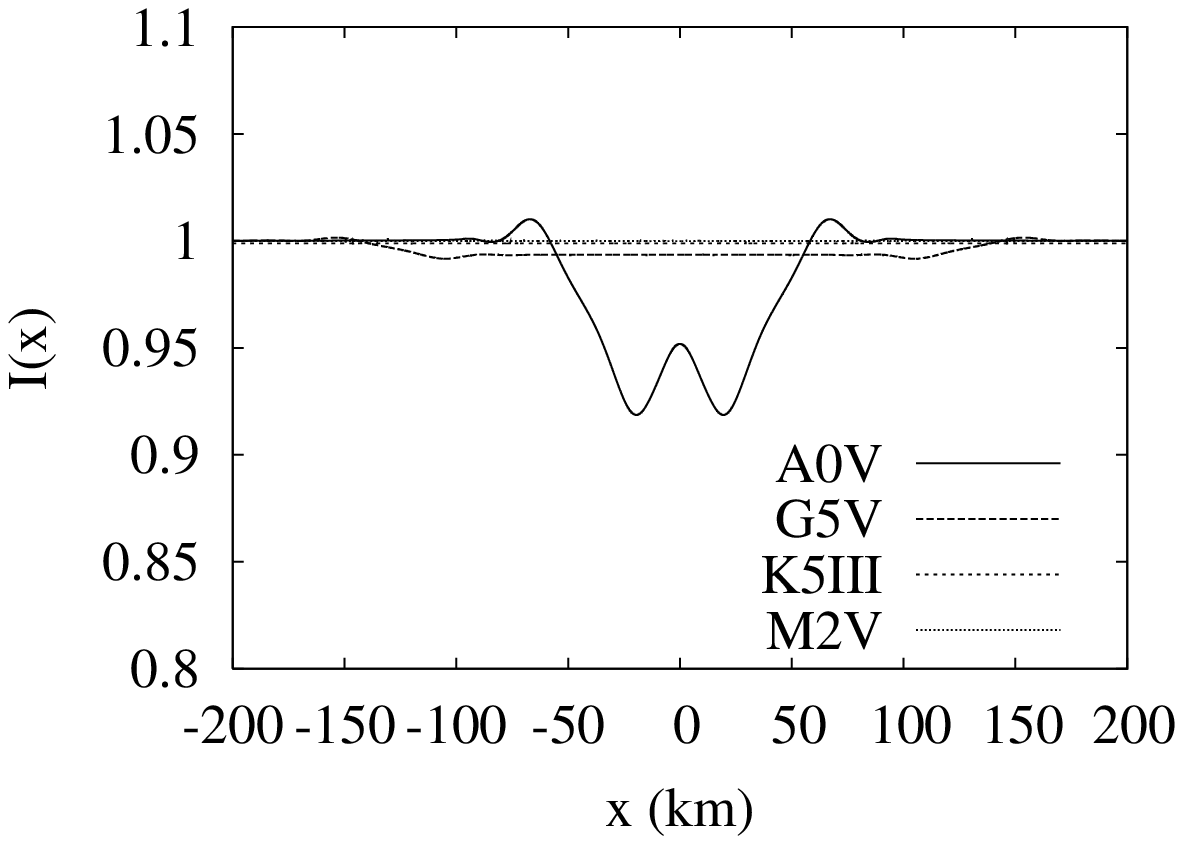}
  \caption{Diffraction profiles for an Oort Cloud object for different
    $V=12$ background stars: A0V, G5V, K5III, and M2V.  The occulting
    object for this set of curves has a radius of 10~km and is at a
    distance of 10,000~AU.  Projected stellar radii at 10,000~AU are
    43.5, 129.4, 306.3, and 641.4~km respectively.}
  \label{fig:finite_1km_10000au}
\end{figure}

The smoothing and broadening effects at distances well beyond the
Kuiper Belt are more drastic as is expected due to the dependence of
the projected source size on the distance $a$.  Occultation patterns
were calculated for four $V=12$ background source stars for a $r=5$~km
Extended Disk object at $a =1,000$~AU
(Figure~\ref{fig:finite_1km_1000au}) and for a $r=10$~km Oort Cloud
object $a=10,000$~AU (Figure~\ref{fig:finite_1km_10000au}).  These
objects are detectable with the A0V star, but they would be extremely
difficult to detect with the K5III or M2V stars because of the very small
occultation depth due to the large projected radius of the source
star.

\begin{figure}
  \epsscale{1.0} \plotone{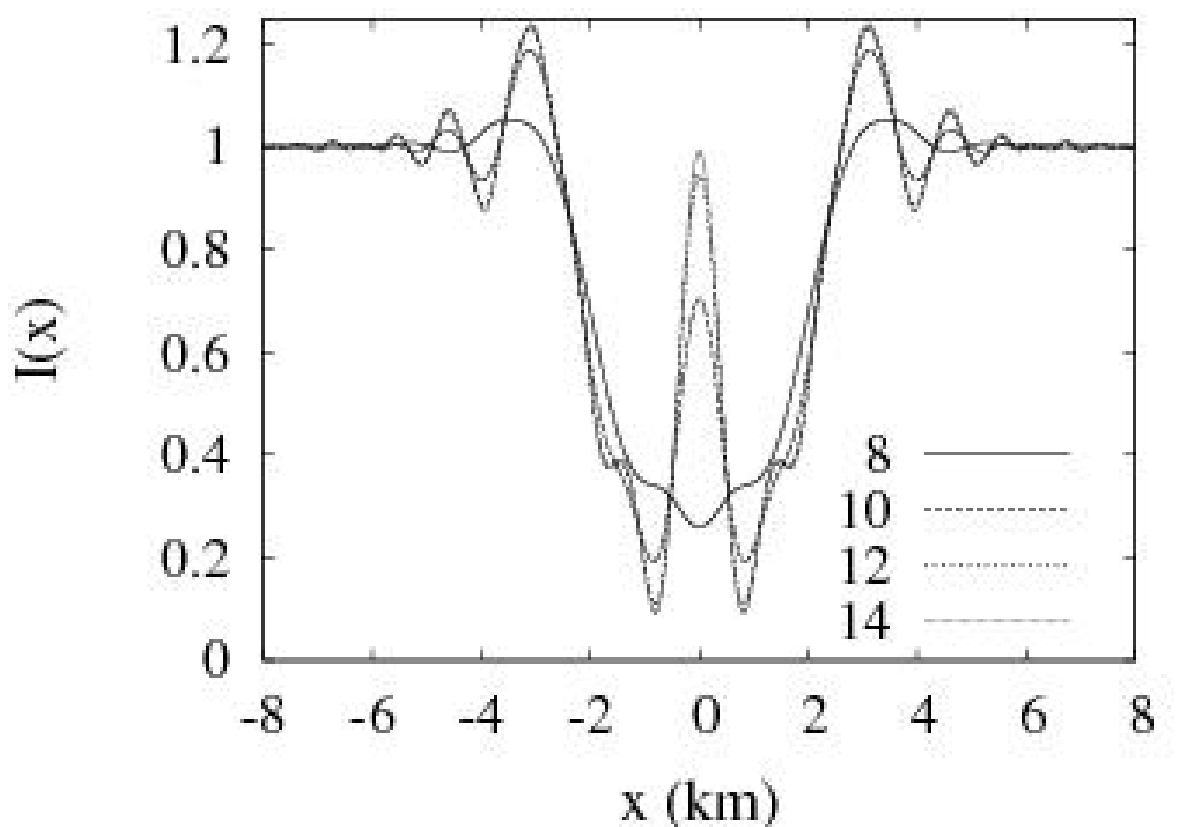}
  \caption{Diffraction profiles for background A0V stars of varying
    magnitudes: $V = 8$, 10, 12, and 14.  The lower right-hand key
    lists V-magnitude values for each curve.  Apparent stellar radii
    at 40~AU are 1.1, 0.44, 0.17, and 0.069~km respectively.}
  \label{fig:mag}
\end{figure}

The angular size of a star depends not only on its spectral type, but
on its magnitude as well.  Figure~\ref{fig:mag} shows the smoothing
effects due to various stellar brightnesses ($V=8$, 10, 12, and 14)
for an A0V star occulted by a $r = 1.5$~km object at 40~AU.  The
stellar radii projected into the plane of the object are 1.1, 0.44,
0.17, and 0.069~km, respectively.  The change in radius from a 14th to
10th magnitude A0V star will dampen the Poisson peak by $\sim30$\%.
However, overall smoothing due to the increase in stellar radius with
brightness is minimal for this range. In the case of a $V=8$ A0V star
which has a projected source radius comparable to both the Fresnel
scale and object radius, the smoothing is significant enough to
eliminate any trace of the Poisson peak and most of the other
diffraction features.

\begin{figure*}
 \epsscale{1.0} \plotone{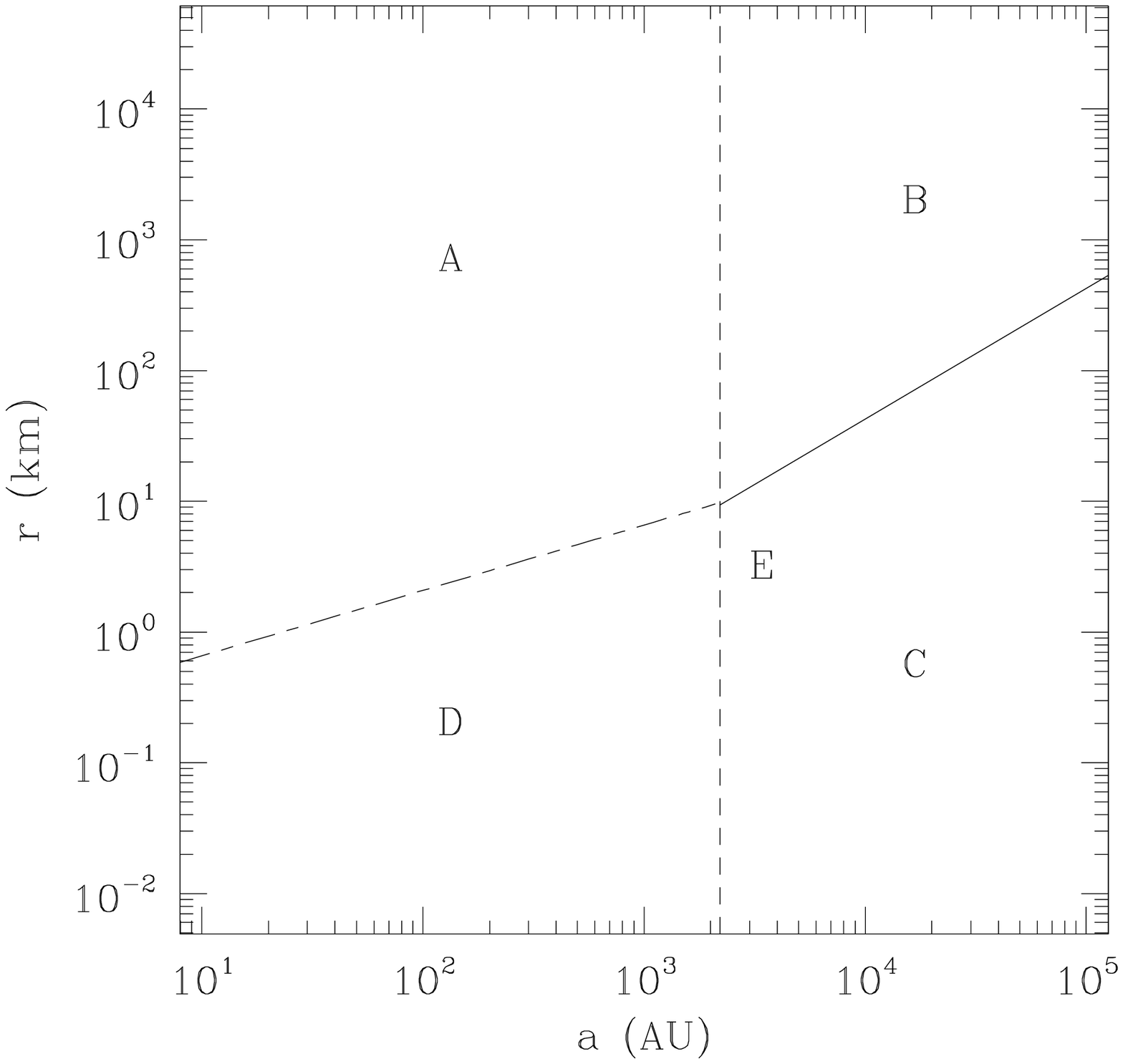}
  \caption{Plot summarizing resulting lightcurve classes for object
    radii $r$ and distances $a$ passing before a $V=12$ A0V star.
    Each class is associated with a region demarcated by the drawn
    lines and is labeled accordingly (A, B, C, D, and E).
    Characteristic curve shapes are dependent upon stellar disk radius
    $r_{\ast}$ (solid line) and the Fresnel scale $F$ (dotted - dashed
    line).  Exemplary curves for each of the listed regions are shown
    in Figure~\protect\ref{fig:regionsb}.}
  \label{fig:regionsa}
\end{figure*}

\begin{figure*}
 \epsscale{1.0} \plotone{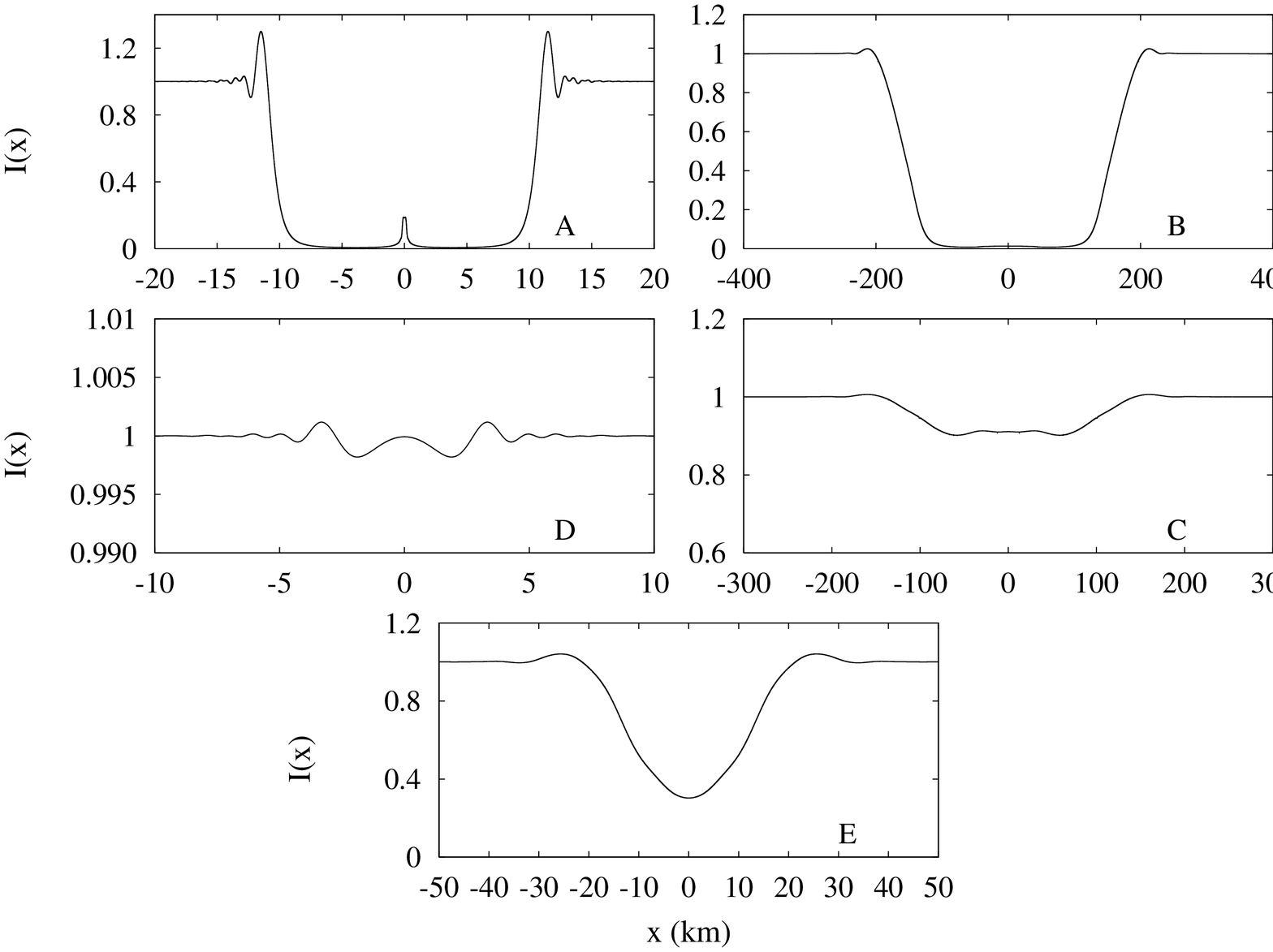}
  \caption{Plots of example lightcurves for each of the five
    morphologies shown in Figure~\protect\ref{fig:regionsb}.  Each
    curve is labeled with it's corresponding region.}
  \label{fig:regionsb}
\end{figure*}

Occultation patterns for stars of various types and brightnesses, and
for objects of various sizes and distances, exhibit a wide range in
morphologies as can be seen from the previously discussed figures.  A
summary of the expected occultation pattern morphologies is presented
in a plot of object radius $r$ versus distance $a$
(Figure~\ref{fig:regionsa}) with accompanying example occultation
patterns (Figure~\ref{fig:regionsb}).  As described above, the
occultation pattern depends upon the projected source size $r_{\ast}$
relative to the Fresnel scale $F$.  The morphologies are then split
into five regions labeled A, B, C, D, and E.  Regions A and D lie in
the distance range $a \lesssim 2,000$~AU where the projected source
radius $\rho_{\ast}<1$. Conversely, regions B and C lie in the
distance range $a \gtrsim 2,000\rm{\ AU}$ where $\rho_{\ast} >
1$. Region E lies in the parameter space $\rho_{\ast} \sim 1$.

Recall that when $\rho \gg 1$, the occultation pattern for a point
source can be approximated by a geometric shadow.  The A and B regions
therefore represent the area of the parameter space for geometric
occultation patterns.  The difference between the two arises from the
source size $\rho_{\ast}$.  At large distances where $\rho_{\ast}
\gtrsim 1$ (region B), significant smoothing by the stellar source
disk washes out all diffraction details including the Poisson peak. To
contrast, diffraction features remain in the geometric occultation
pattern when $\rho_{\ast} \lesssim 1$ (region A).

Occultation patterns which fall into regions C and D in
Figure~\ref{fig:regionsa} fall within the Fresnel and Fraunhofer
regimes ($\rho\lesssim 1$).  They both exhibit shallower occultation
depths, but again, depending on the source size $\rho_\ast$, the
occultation will exhibit differing occultation width broadening and
smoothing. In region C, the Fresnel scale is smaller than the
projected source size, and therefore only a fraction of the stellar
flux is diffracted. These events are wide but extremely shallow.
Occultation patterns in Region D simply resemble the Fraunhofer
patterns we discussed previously because the small source sizes here
have little effect.  Finally, patterns in region E, where $\rho_\ast
\sim 1$ and $\rho \sim 1$ , show a significant depth without complete
extinction of the source, and are devoid of significant diffraction
features.

From the above discussion, it follows that in order to minimize
smoothing and maximize the occultation depths, selected target stars
should be relatively dim and blue, since brighter and redder stars
have larger apparent disks. With this choice a survey would have to
contend with reduced signal-to-noise.  With better photometry,
improved signal-to-noise augmented by brighter background sources may
slightly increase the detection of smaller objects as the occultation
event widths are extended by the stellar disk (see
Equations~\ref{eq:width_fsu_finite}~and~\ref{eq:width_finite}).  This
approach however, would sacrifice diffraction details and diminish the
depths of occultation events.  Background source selection is
furthermore complicated by the linear dependence of the projected
source radius $r_{\ast}$ on $a$.  Effective selection of target stars
therefore depends on considerations such as the target population,
observational parameters, and expected event rates, as well as
lightcurve shape.

\section{Relative Velocity and Sampling}\label{sec:velocity}
Rather than observing stationary occultation patterns like the ones
that have so far been discussed, real surveys will observe these
patterns in time due to the relative velocity of the object transverse
to the line of sight. The main source of the transverse motion is the
orbital motion of the Earth, with a small contribution from the
velocity of the occulting object itself.  Assuming that the occulting
object lies in the ecliptic plane on a circular orbit the transverse
velocity expression is
\begin{equation}
  v_\mathrm{T} = v_\mathrm{E}\left(\cos\phi -
 \sqrt{\frac{a_\mathrm{E}}{a}
   \left(1-\frac{a^2_\mathrm{E}}{a^2}\sin^2\phi\right) }\right),
\label{eq:vel_elong}
\end{equation}
where $\phi$ is the angle of opposition, the angle between the object
and the opposing direction of the sun, $v_\mathrm{E}=
29.8$~km~s$^{-1}$, is the orbital speed of Earth, and $a_\mathrm{E} =
1$~AU is the distance from the Sun to the Earth.  At opposition ($\phi
= 0^\circ$) where the transverse velocity reaches a maximum, the
expression for $v_\mathrm{T}$ simplifies to
\begin{equation}
  v_\mathrm{T} = v_\mathrm{E} \left ( 1 - \sqrt{\frac{1\rm{\ AU}}{a}} \right).
\label{eq:velocity}
\end{equation}
A typical KBO at a distance of $a=40$~AU will have a transverse
velocity $v_\mathrm{T}\approx 25$~km~s$^{-1}$.  The occultation
profile is then measured as a lightcurve in time
$t=x/v_\mathrm{T}$. Given the occultation width (see
Equation~\ref{eq:width_finite}), the occultation has a duration
\begin{equation}\label{eq:duration}
T = \frac{W^{\ast}}{v_\mathrm{T}}. 
\end{equation}

A true survey will collect images by temporally integrating
infinitesimally sampled lightcurves over a finite exposure time
$\Delta t$.  Shutter speed, frame transfer, and other photometric
delays will lead to time lags between images. However, here we assume
that any delays that arise from the photometric system are minimal and
that the exposure time is equal to the inverse of the sampling
frequency $f = 1/\Delta t$.  The discrete lightcurve is then given by
points integrated in time
\begin{equation}\label{eq:discrete_intensity}
I(t_{i}) = \int\limits_{t_{i}-\Delta t/2}^{t_{i}+\Delta t/2}I(t)dt.
\label{eq:mphot}
\end{equation}

To give an example of how an occultation pattern is measured,
accounting for finite exposure, consider observations made at
opposition with relatively fast $f=20$~Hz sampling. This would
correspond to an exposure time $\Delta t\approx50$~ms. For a KBO at
40~AU, intensity values will be integrated over a distance of $\Delta
x\approx1.3$~km.  This is near the target KBO size of TAOS, and near the
Fresnel scale for KBOs and objects in the Extended Disk.  At Oort
cloud distances the integration distance $\Delta x$ is roughly ten
times smaller than the Fresnel scale.

\begin{figure*}
 \epsscale{1.0} \plotone{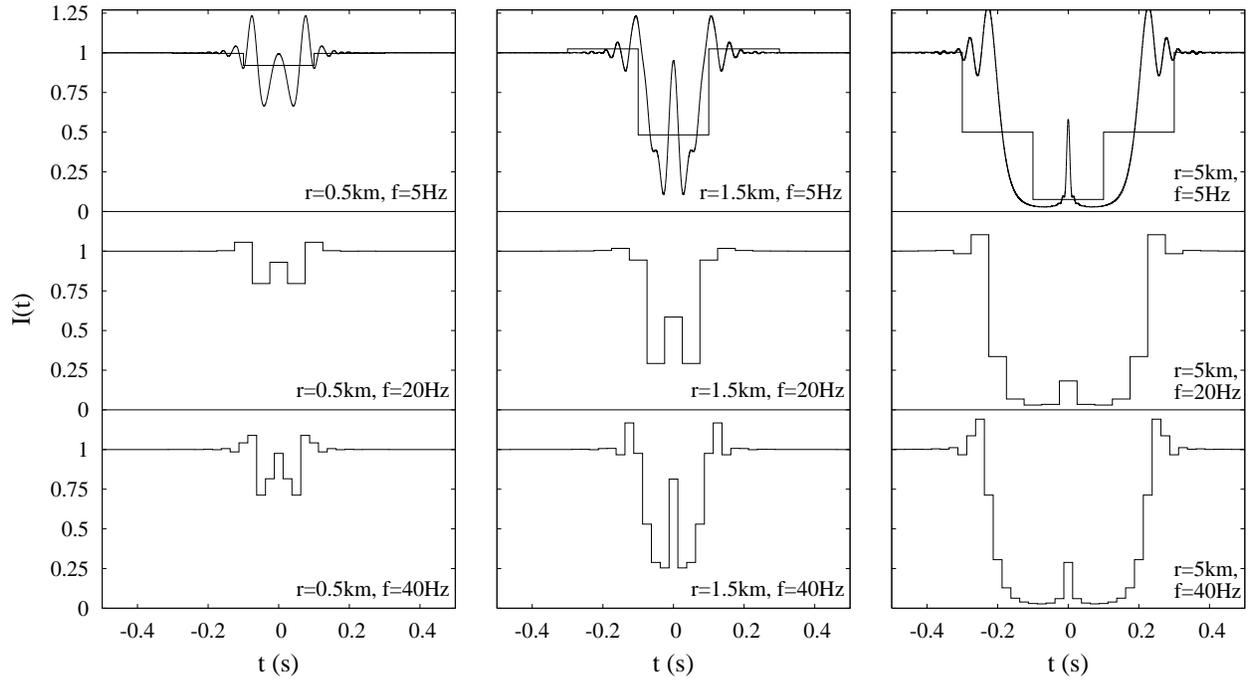}
  \caption{Diffraction profiles observed at opposition for a $V=12$
  A0V background star occulted by $r=0.5$, 1.5, and 5~km objects at
  40~AU.  Each curve is continuously sampled at one of three
  frequencies: 5, 20, and 40~Hz.  The unsampled lightcurve for each
  object are shown in the top panels.}
  \label{fig:smp_40au}
\end{figure*}

\begin{figure*}
 \epsscale{1.0} \plotone{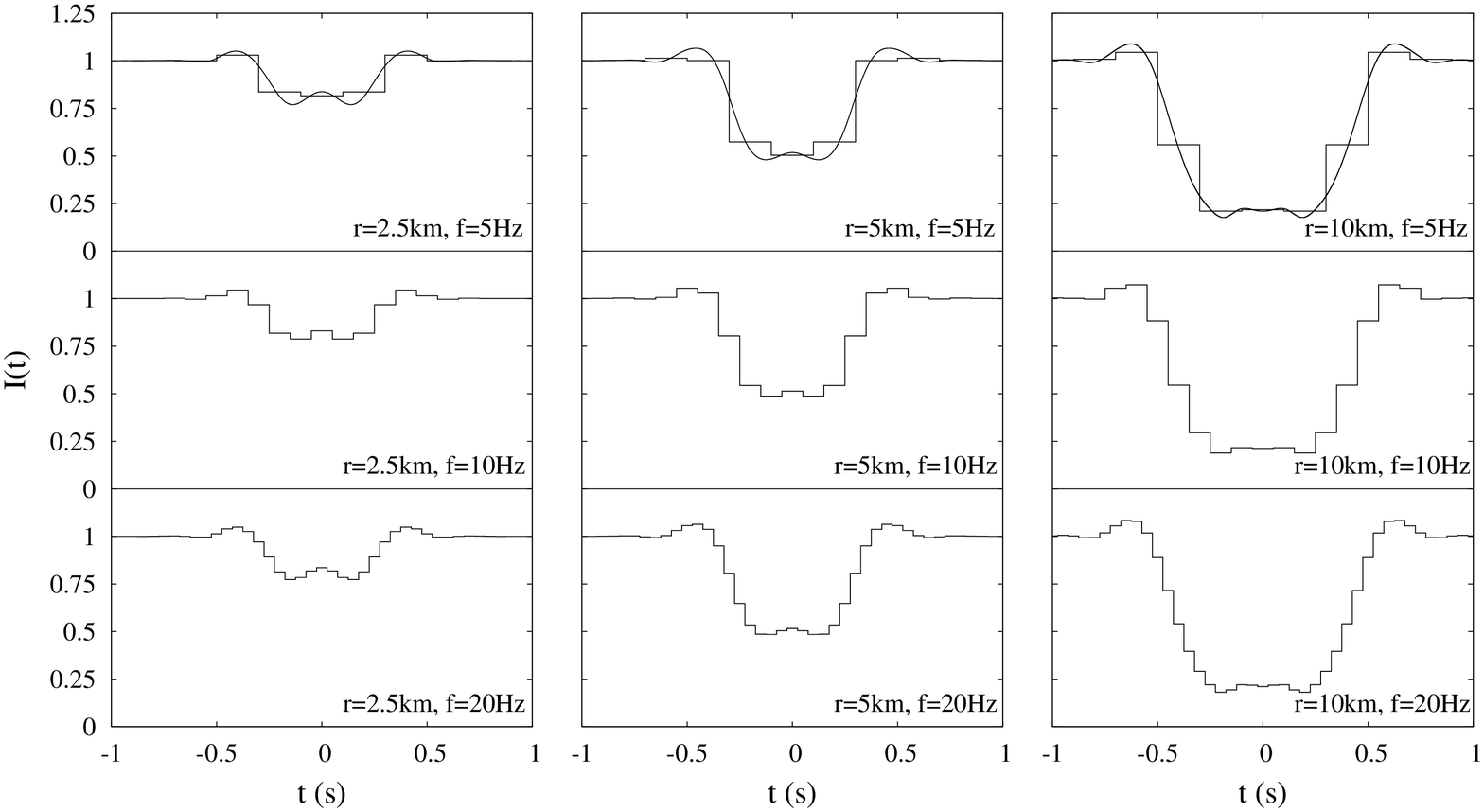}
  \caption{Diffraction profiles observed at opposition for a $V=12$
  A0V background star occulted by $r=2.5$, 5, and 10~km objects at
  1,000~AU.  Each curve is continuously sampled at one of three
  frequencies: 5, 10, and 20~Hz.  The unsampled lightcurve for each
  object are shown in the top panels.}
  \label{fig:smp_1000au}
\end{figure*}

\begin{figure*}
 \epsscale{1.0} \plotone{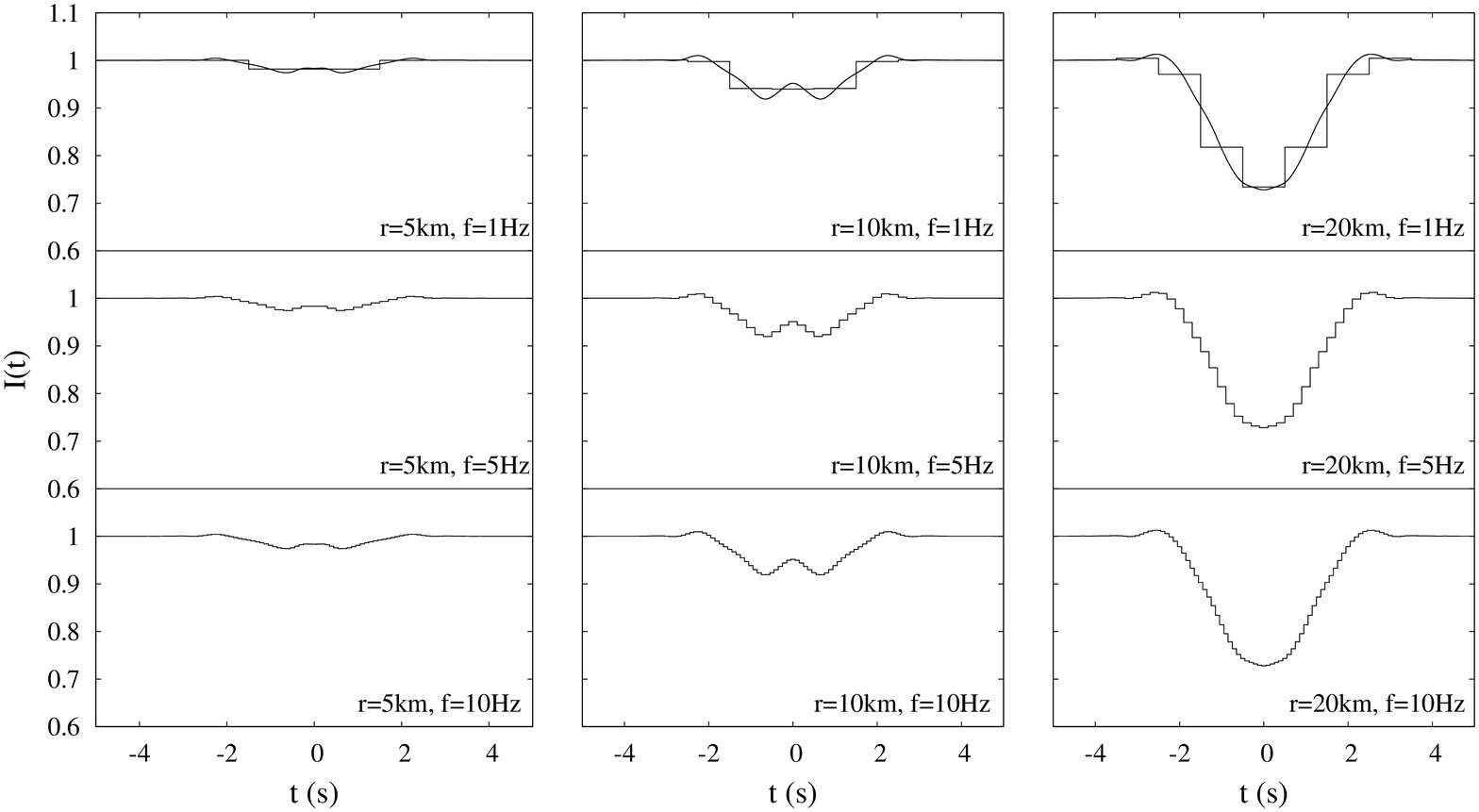}
  \caption{Diffraction profiles observed at opposition for a $V=12$
  A0V background star occulted by $r=5$, 10, and 20 km objects at
  10,000~AU.  Each curve is continuously sampled at one of three
  frequencies: 1, 5, and 10~Hz. The unsampled lightcurve for each
  object are shown in the top panels.}
  \label{fig:smp_10000au}
\end{figure*}

The sampling rate is a critical parameter for an occultation survey. 
In order to resolve diffraction effects, the sampling rate of a survey
must be high enough such that $\Delta x \ll W^\ast$.  In cases where
$\Delta x \sim W^\ast$, events will still be detectable however,
diffraction effects will not be evident in the lightcurve because the
entire event will appear in only one or two samples.  However, when
$\Delta x \gg W^\ast$ events become difficult to detect because all
the power of the diffraction occurs within a small fraction of a
single sample time and is therefore averaged out. This is illustrated
in Figures~\ref{fig:smp_40au}, \ref{fig:smp_1000au},
and~\ref{fig:smp_10000au}.

Figure~\ref{fig:smp_40au} depicts occultation pattern profiles for $r
=$~0.5, 1.5, and 5~km objects at 40~AU occulting an A0V $V=12$~star
sampled at frequencies of 5, 20, and 40~Hz.  A sampling of 5~Hz
corresponds to $\Delta x\approx5\rm$~km at 40~AU, which is comparable
to the minimum event width in the Fraunhofer regime.  It can
be seen that for the $r=0.5$ and 1.5 km objects, only one data point
shows significant deviation from the otherwise flat lightcurve.  On
the other hand, for the $r=5$ km object, the width is slightly larger
than the sampling size, and thus three data points are affected.  In
all three cases, diffraction effects cannot be discerned in the
lightcurves.  In the case of 20~Hz sampling, the widths of the $r=0.5$
and 1.5~km events are slightly larger than the sampling distance
$\Delta x$ and thus typically three data points are affected.  For the
$r=5$~km event, the sample size is small enough relative to the event
width that diffraction effects are slightly visible.  Finally, in the
case of the 40~Hz sampling, $\Delta x \ll W^\ast$, diffraction effects
can be seen for all three objects.

Figures~\ref{fig:smp_1000au}~\&~\ref{fig:smp_10000au} show that at
larger distances, the occultation width is augmented by the size of
the Fresnel scale and size of the source star and the sampling
requirements are therefore less severe.  Due to the large relative
size of $r_\ast$ most of the diffraction effects are washed out anyway
so no gain can be obtained from higher sampling.  In fact, because the
the occultation depths are significantly shallower at larger distances,
signal-to-noise considerations will require the lowest sampling rate
possible that will still allow the event to be resolved.

Note that observations could be made away from opposition, which would
have the effect of increasing the duration of an event. This would
allow for longer sample times and increased signal-to-noise for each
measurement. However, this comes at a cost to the overall event rate
\citep{CandF2003}. To simplify the discussions throughout the remainder
of this paper, it will be assumed that all measurements are made at
opposition. The effects of observing away from opposition will be
revisited in \S~\ref{sec:detect} and \S~\ref{sec:threshold}.

\section{Noise}\label{sec:noise}
In order to determine the observable target population of an
occultation survey, it is necessary to consider the effects of noise.
Three photometric systems, TAOS, Whipple, and the MMT, are described
here, and estimates of signal-to-noise for each system are used to
simulate occultation curves.  In the discussion that follows, $S$ is
used to indicate the signal-to-noise ratio of a $V=12$ A0V star. 

TAOS utilizes three to four $50$~cm wide-field telescopes installed at
the Lu-Lin Observatory in central Taiwan.  Details of the survey can
be found in several references \citep[e.g.][]{Lehner2006, King2002}.
Simultaneous monitoring for occultations on multiple telescopes allow
for the rejection of false positive events.  Each telescope is
equipped with a $2048\times2048$ CCD array with a $3~\Box^\circ$ field
of view.  The TAOS cameras read out at $5$~Hz and monitor anywhere
from several hundred to a few thousand stars in a target field.  The
reported signal-to-noise for an A0V $V=12$ star is
$S_{\mathrm{TAOS}}\sim 8$ \citep{Lehner2007}.

A campaign on the 6.5~m Multiple-Mirror Telescope (MMT), located at
the Whipple observatory on the summit of Mount Hopkins, Arizona, has
been proposed \citep{Bianco2007}. The survey would utilize Megacam, an
array of 36 edge-butting $1024\times2304$ pixel CCDs.  Megacam would
be used in continuous readout mode, achieving a sampling rate of
200~Hz (in this paper we assume a sampling rate of 20~Hz as we expect
the observations to be binned) on $0.16~\Box^\circ$ fields. For this
system the monitoring of several hundred bright stars ($\sim 500$ with
magnitude $V\leq 15$) is possible, and test images show a
signal-to-noise ratio of $S_\mathrm{MMT} \geq 70$ for stars with
magnitudes $V\leq13$ can be achieved.  This would lead to the
collection of $\sim 10^4$ star hours over 3 nights.

A dedicated telescope system like that of TAOS with the ability to
sample lightcurves with high signal-to-noise and fast sampling rates
as the previously described campaign on the MMT, is the ideal system.
Whipple, a proposed space telescope dedicated to the detection of
occultations by Outer Solar System objects \citep{Kaplan2003}, is an
attempt to match these requirements. It is based on the design of the
Kepler Mission \citep{Koch2005}, with the same a $100~\Box^\circ$ field of
view, but has a modified focal plane designed to allow read out
cadences of up to $40$~Hz.  For the purposes of this discussion a
signal-to-noise ratio of $S_{\mathrm{Whipple}}\sim 40$ is assumed.
This accounts for both Poisson statistics and read noise
\citep{Lehner2007}. A dedicated space telescope has several
advantages over ground based campaigns such as those mentioned above.
Dedicated monitoring in space means that continuous monitoring of
fields are a possibility with larger fields of view.  It is estimated
that Whipple will observe up to $\sim140,000$ stars in a given field.
Hence, the larger number of monitored stars increases the rate of
detection which would provide more sound statistics on the surface
density of small objects. A space-based telescope also has the
advantage that it is not affected by atmospheric scintillation, which
would allow the survey to be sensitive to objects smaller and more
distant to those that could be observer from the ground.

\begin{figure*}
 \epsscale{1.0} \plotone{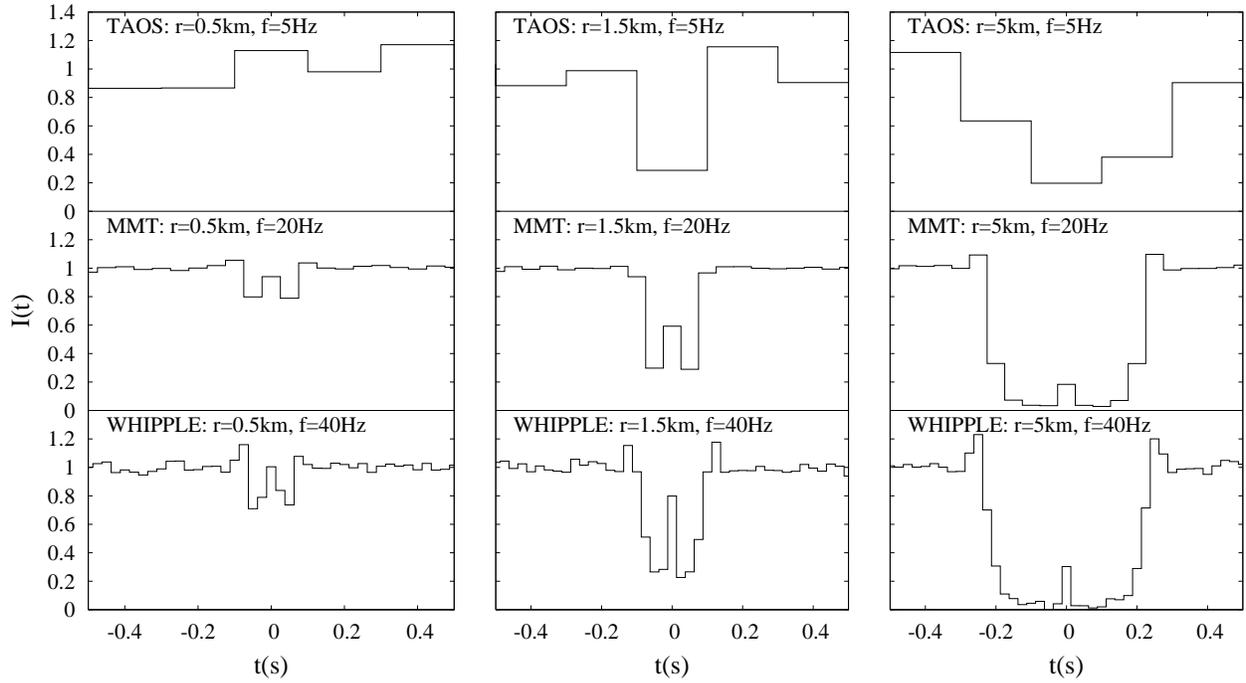}
  \caption{Diffraction profiles with added noise observed at
    opposition for a $V=12$ A0V star occulted by objects with $r =
    0.5$, 1.5, and 5~km at 40~AU. MMT and Whipple lightcurves are
    simulated with signal-to-noise ratios of 70 and 40 respectively,
    and TAOS lightcurves are generated with a signal-to-noise ratio of
    8.}
  \label{fig:nse_40au}
\end{figure*}

For the Kuiper Belt, TAOS's modest 5~Hz sampling and signal-to-noise
of $8$ is sufficient for the detection of objects with
$r\gtrsim1.5$~km, but provide none of the diffraction details
(Figure~\ref{fig:nse_40au}).  The MMT and Whipple photometry have
greater signal-to-noise and much higher sampling, and therefore are
not only capable of seeing objects smaller than $r=1.5\rm{\ km}$, but
are also able to detect diffraction effects for these smaller objects.

\begin{figure*}
 \epsscale{1.0} \plotone{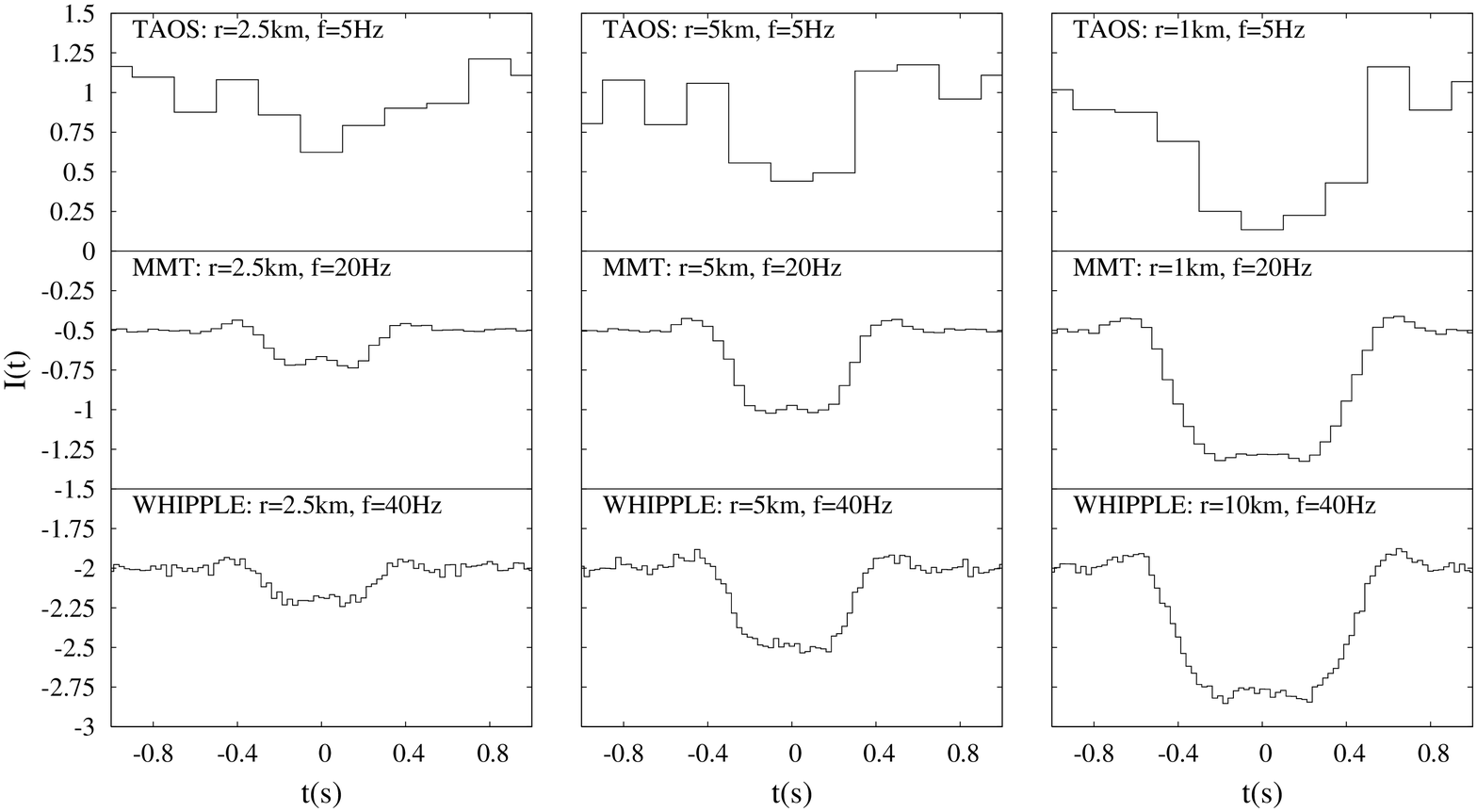}
  \caption{Diffraction profiles with added noise observed at
    opposition for a $V=12$ A0V background star occulted by objects
    with $r = 2.5$, 5, and 10~km at 1,000~AU. MMT and Whipple
    lightcurves are simulated with signal-to-noise ratios of 70 and 40
    respectively, and TAOS lightcurves are generated with a
    signal-to-noise ratio of 8.}
  \label{fig:nse_1000au}
\end{figure*}

Simulations for Extended Disk objects in the region intermediate to
the Kuiper Belt and Oort Cloud ($\sim 1,000~\mathrm{AU}$) lead to a
similar conclusion (see Figure~\ref{fig:nse_1000au}).  Although not
the stated focus of TAOS, it turns out that TAOS could potentially
provide constraints on the number and size of Sedna's smaller cousins,
as objects with $r\sim 5$~km are potentially detectable with the TAOS
photometry.  Objects much smaller in size would unlikely be seen by
TAOS. The higher signal-to-noise of an MMT-based survey and the
proposed Whipple Space Telescope, coupled with rapid sampling make the
detection of the weak diffraction effects possible.

\begin{figure*}
 \epsscale{1.0} 
  \plotone{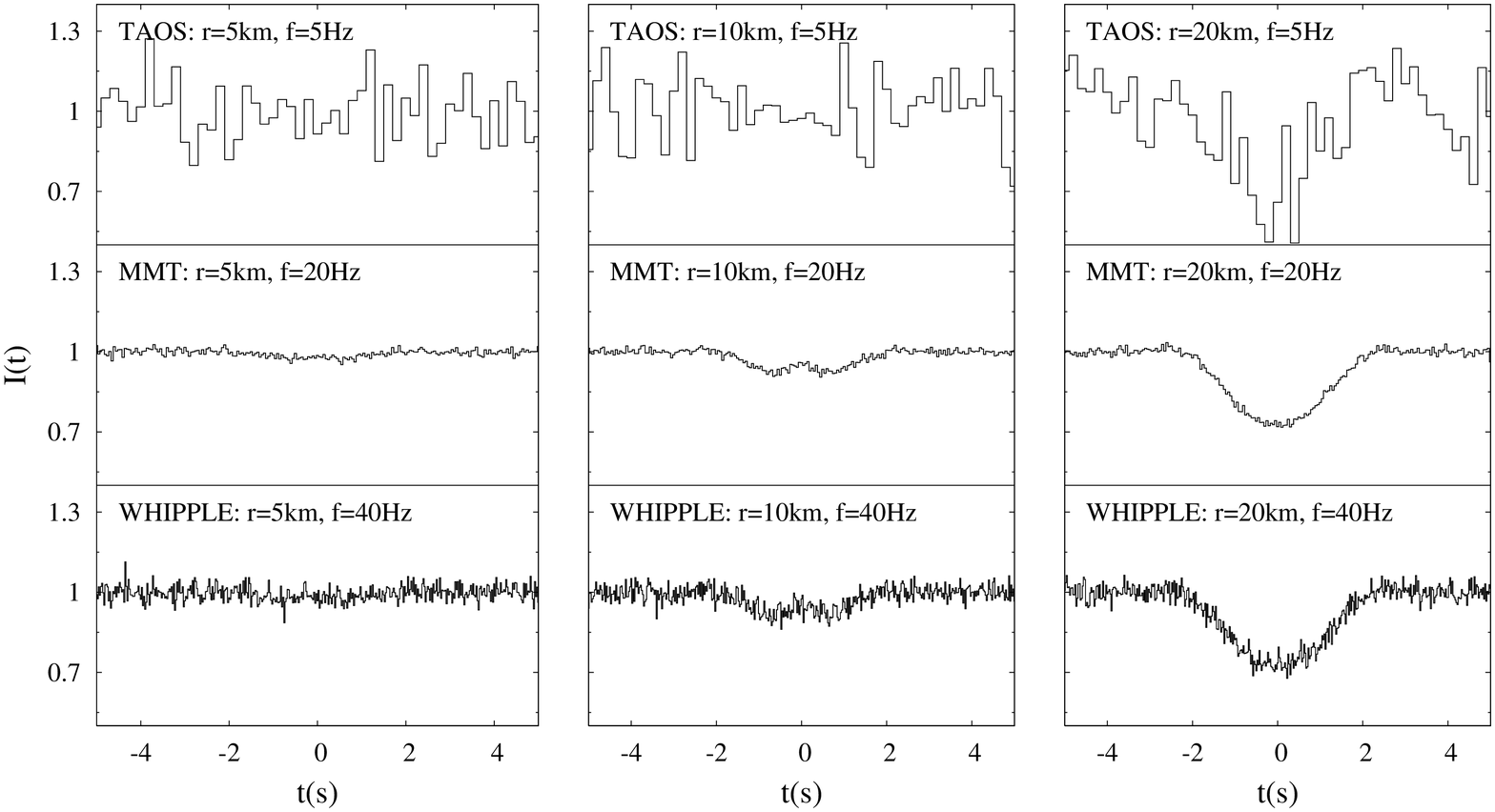}
  \caption{Diffraction profiles with added noise observed at
    opposition for a $V=12$ A0V background star occulted by objects
    with $r = 5$, 10, and 20~km at 10,000~AU. MMT and Whipple
    lightcurves are simulated with signal-to-noise ratios of 70 and 40
    respectively, and TAOS lightcurves are generated with a
    signal-to-noise ratio of 8.}
  \label{fig:nse_10000au}
\end{figure*}

The top panels of Figure~\ref{fig:nse_10000au} indicate that small
objects ($r\sim 10$~km) in the Oort Cloud region will not provide
enough signal for TAOS. Events from a $r = 10$~km object at
$10,000$~AU is barely detectable by the MMT and Whipple surveys, and
larger objects upwards of $r = 20$~km in size are a possibility for
all three surveys.  At distances $a \gtrsim 1,000$~AU, the difference
between $20$ and $40$~Hz sampling is minimal.  Both of these sampling
frequencies, at large distances where the event durations are longer,
are sufficient to determine the shape of the occultation
profile. However, it is clear from Figure~\ref{fig:nse_40au} that
there is a clear benefit of sampling at $40$~Hz versus $20$~Hz when
searching for events from KBOs at 40~AU.

\section{Detectability}
\label{sec:detect}
It would be useful to describe an occultation event by a single
parameter which can be used in conjunction with the parameters of a
survey to determine whether or not the occultation event has
sufficient signal to be detected above the underlying noise of a
measured lightcurve. For ideal curves parameterized by the Fresnel
distance scale $\eta$ and radius $\rho$, we thus define the
\emph{detectability} parameter as
\begin{equation}\label{eq:detect_int_fsu}
\Xi_{\rm{FSU}} = \int_{-\infty}^{\infty}( I_{\rho}(\eta) - 1 )^{2} d\eta 
\end{equation}
When curves are discretely sampled, the detectability is a sum over a
discrete set of observations ${I_{\rho}(\eta_{i}})$.  If $m$ samples
are taken over intervals of $\Delta\eta$ corresponding to a sampling
frequency $1/\Delta\eta$, then
\begin{equation}\label{eq:detect_sum_fsu}
\Xi_{\rm{FSU}} = \sum_{i=1}^{m}( I_{\rho}(\eta_{i}) - 1 )^{2}
\Delta\eta
\end{equation}
Note that for a geometric occultation where $\rho \gg 1$, complete
extinction ($\Delta = 1$) of the background source throughout the
duration of the occultation event means that the detectability of a
geometric occultation is roughly equal to the occultation width
($\Xi_{\rm{FSU}} \sim \Omega$).  Lightcurves in this regime clearly
have the largest detectability, while the shallow depths for
Fraunhofer occultation events makes such events the least detectable
of the three regimes.

\begin{figure}
  \epsscale{1.0} \plotone{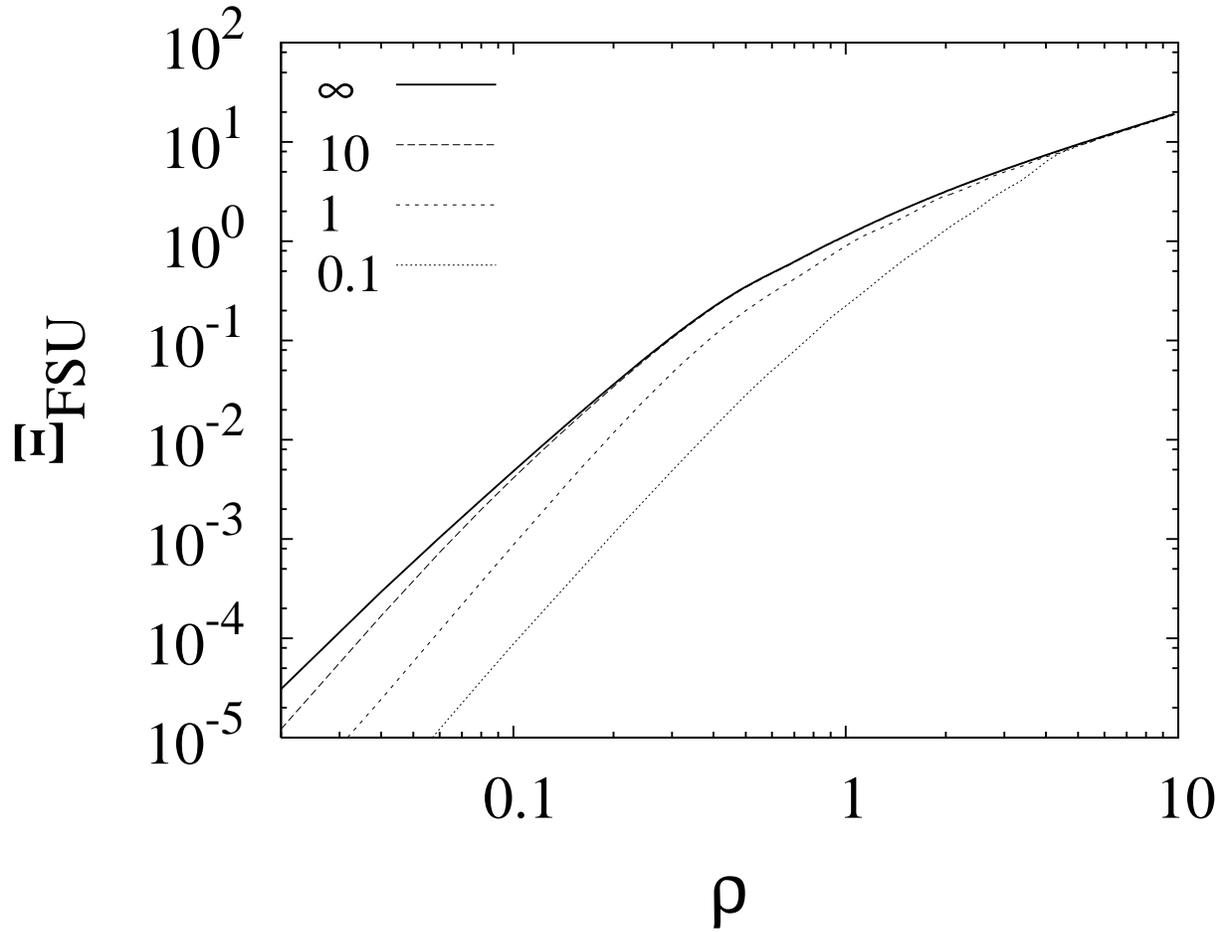}
  \caption{Dimensionless detectability (for a point background source)
    plotted versus the occulting object radius in Fresnel scale units
    at various sampling rates $1/\Delta \eta$ listed in the left-hand
    key.  Low sampling rates result in lower detectability at smaller
    values of $\rho$.}
  \label{fig:detect_fsu}
\end{figure}

Figure~\ref{fig:detect_fsu} is a plot of $\Xi_{\rm{FSU}}$ versus
$\rho$.  The solid line traces the detectability for the ideal
continuous occultation intensity profile as expressed in
Equation~\ref{eq:detect_int_fsu}.  At $\rho\gtrsim10$, the
detectability asymptotes to the occultation width $2\rho$.  The effect
of finite sampling is also shown in this plot for three sampling
frequencies $1/\Delta\eta =10\rm{,\ } 1\rm{,\ and\ } 0.1$.  Lower
sampling rates result in a decreased ability to detect occultation
events.  For larger values of $\rho$, this effect is diminished as
occultation event widths increase.  This is because larger objects
produce wide geometric occultation patterns which completely
extinguish the background source and therefore detection of objects
with larger $\rho$ is not greatly improved with increases in the
sampling frequency.

The scaled detectability $\Xi_{\rm{FSU}}$ shows the general trend for
all occultation patterns scaled to the Fresnel scale.  However, it
provides little intuition for event detection in a real survey.  The
above discussion is extended to the physical parameter space of $a$
and $r$ by introducing the corresponding expression in physical units to
Equation~\ref{eq:detect_int_fsu}:
\begin{equation}\label{eq:detect_int}
\Xi = \int_{-\infty}^{\infty}( I_{r,a}(t) - 1 )^{2} dt.
\end{equation}
Note that $\Xi$ now has the dimension of time, and is related to
$\Xi_\mathrm{FSU}$ as
\begin{displaymath}
\Xi=\Xi_\mathrm{FSU}F/v_\mathrm{T}.
\end{displaymath}

\begin{figure}
  \epsscale{1.0} \plotone{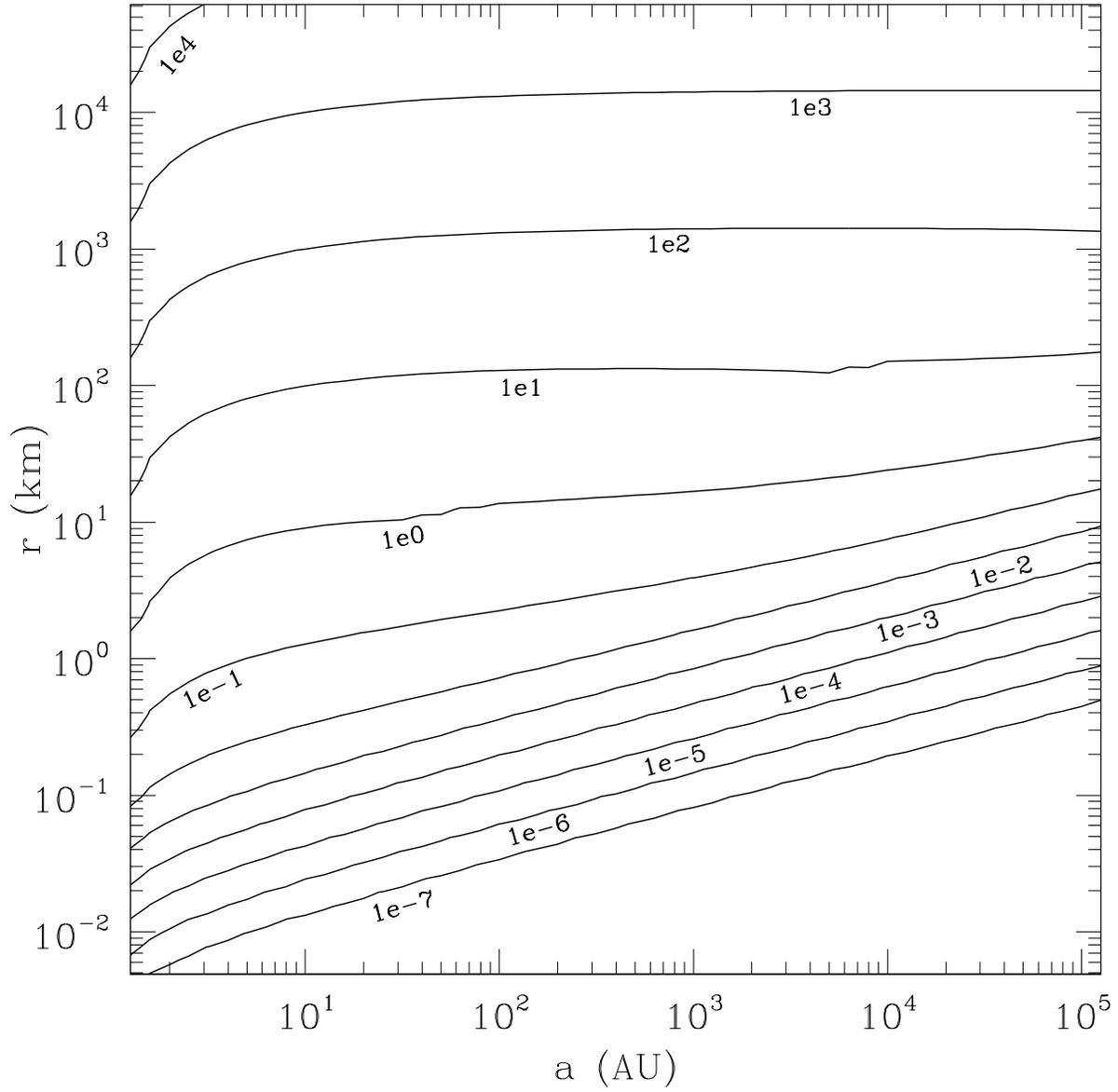}
  \caption{Detectability contour plot of objects as a function of
    diameter and distance for a point source.  An A0V spectrum was
    assumed.}\label{fig:detect_pts}a
\end{figure}

\begin{figure*}
 \epsscale{1.0} \plotone{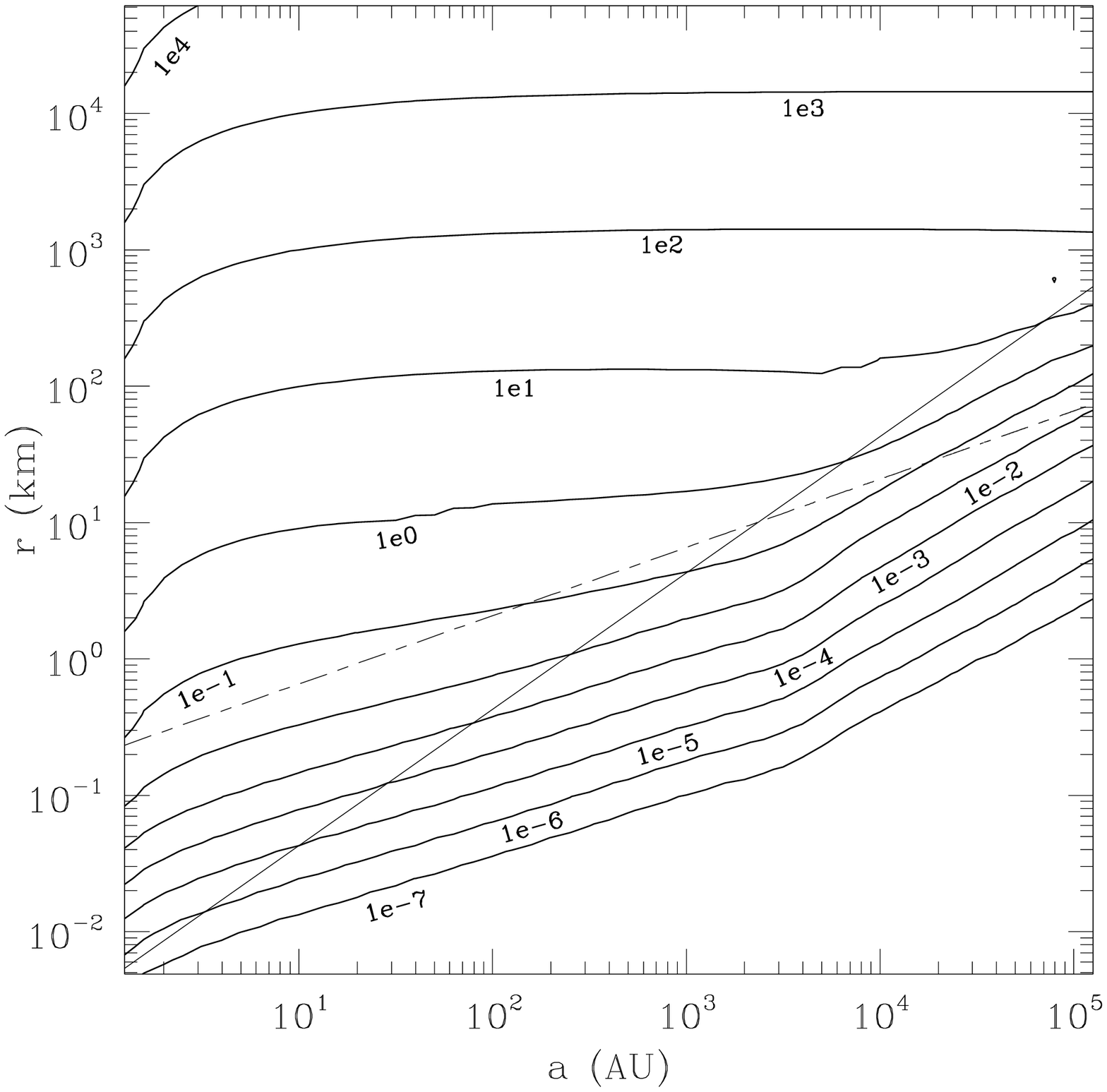}
  \caption{Detectability contour plot of objects as a function of
    diameter and distance for an A0V $V=12$ star.  The finite size of
    the stellar disk was accounted for.  Also shown are the limits
    where the object radius is equivalent to the Fresnel scale $F$
    (dashed-dotted line) and equivalent to the stellar disk (straight
    solid-line).}
  \label{fig:detect_fns}
\end{figure*}

Figure~\ref{fig:detect_pts} is a contour plot of the detectability for
occultations of an A0V point source as a function of $a$ and $r$.  The
detectability increases for objects at closer distances and decreases
at larger distances.  It can also be seen that larger objects show
less drastic changes in the detectability as their distance is
increased or decreased.  Similarly, Figure~\ref{fig:detect_fns} shows
a detectability contour plot for occultations of a finite size V=12
A0V star by objects at various distances $a$ and with different radii
$r$. Note that the two plots are similar at lower values of $a$ and
larger values of $r$, but for smaller objects at larger distances,
when the projected source size becomes larger than both the Fresnel
scale and the object size, the detectability of the event drops. The
upturn in the contours of Figure~\ref{fig:detect_fns} roughly
correspond to the distance $a$ where the projected source size is
equal to the Fresnel scale. Note that for dimmer stars with smaller
radii, this upturn in the contours will occur at larger distances, and
for an A0V star with magnitude 16 or greater, the detectability
contour plot would be virtually identical to the plot shown in
Figure~\ref{fig:detect_pts}.

\begin{figure*}
 \epsscale{1.0} \plotone{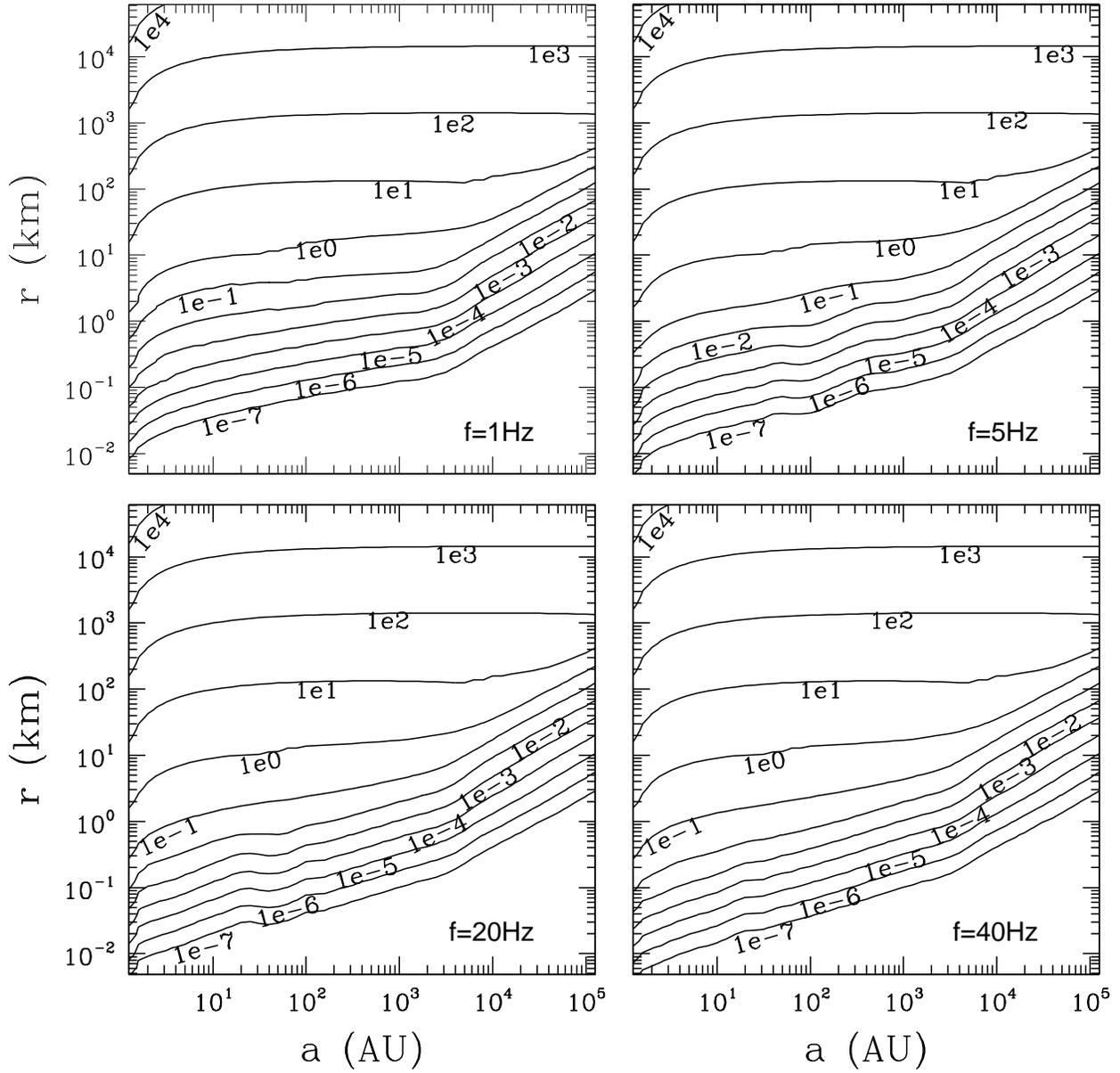}
  \caption{Contour plots of the detectability as a function of radius
    $r$ and distance $a$ for occultation curves sampled at frequencies
    of 1, 5, 20, and 40~Hz.  An A0V $V=12$ star is assumed.}
  \label{fig:detect_smp}
\end{figure*}

As before, a continuously sampled curve at a frequency $f$ and
corresponding exposure $\Delta t$ is a set of discrete observations
$I(t_{i})$, In this case, Equation~\ref{eq:detect_int} becomes
\begin{equation}\label{eq:detect_sum}
\Xi = \sum_{i=1}^{m}{ \Delta t ( I(t_{i}) - 1 )^{2} },
\end{equation}
where $m$ is set to be large enough to accommodate the entire width of
the event.
  The effects of sampling can be seen in
Figure~\ref{fig:detect_smp} which shows contour detectability plots
for a finite-sized A0V $V=12$ star for various occultation curves
sampled at frequencies of 1, 5, 20, and 40~Hz.  Just as in
Figure~\ref{fig:detect_fsu} the detectability changes very little for
geometric occultation light curves irrespective of the sampling.
These curves correspond roughly to detectabilities that are of order 1
or greater.  For regions of the plot where the object radius is either
of order or less than the Fresnel scale, sampling rates will change
the detectability more drastically.  As an example, recall from
Figure~\ref{fig:smp_40au} that a $r=0.5$~km object at $a=40$~AU
generates a fluctuation of $\sim20$\%.  This event sampled at a rate
of $5$~Hz has a detectability of order $10^{-3}$.  The detectability
of this event however is improved ten-fold when observed at $40$~Hz.

\begin{figure}
  \epsscale{0.9} \plotone{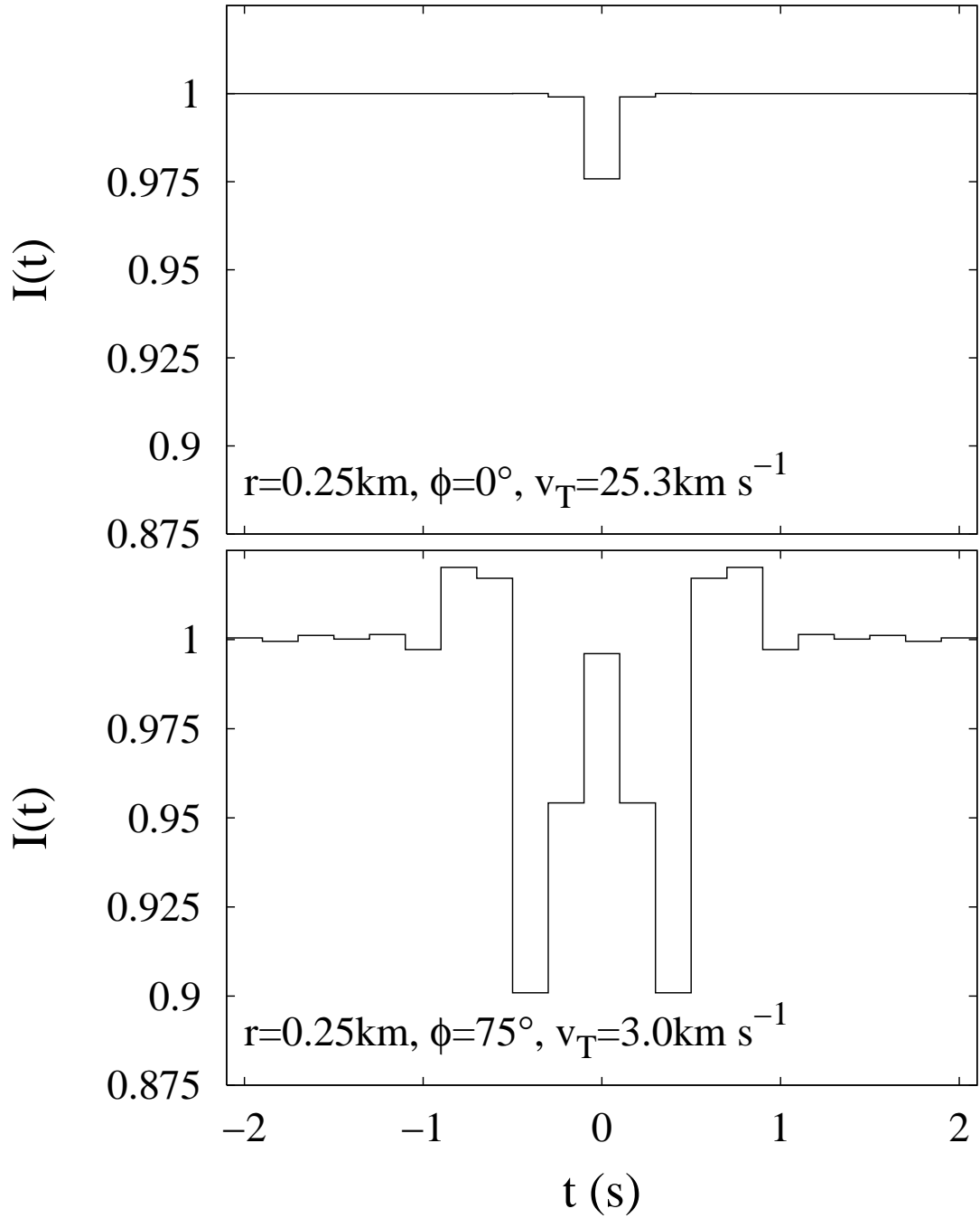}
  \caption{Lightcurves for a $r = 0.5$~km KBO at 40~AU measured at
opposition and at an opposition angle of $75^\circ$. Moving to $\phi =
75^\circ$ reduces the relative velocity $v_\mathrm{T}$ by roughly a
factor of eight, significantly increasing the detectability of the
event.}
  \label{fig:oppkbo} 
\end{figure}

As discussed in \S~\ref{sec:velocity}, observations can be made away
from opposition, reducing the relative velocity of the target
object. This affects the detectability as well. To illustrate,
consider the case of a KBO occultation event measured at $75^\circ$
from opposition. From Equation~\ref{eq:vel_elong}, it can be shown
that the relative velocity $v_\mathrm{T}$ is reduced by a factor of
about eight. If the sampling rate is high enough that the diffraction
effects can be resolved, the event duration $T$, and thus the
detectability $\Xi$, are subsequently increased by a factor of
eight. As mentioned in \S~\ref{sec:velocity}, the event rate for such
events is decreased by a factor of eight as well. In cases where
$v_\mathrm{T}\Delta t \gg W^\ast$ and the event is averaged out with
the nominal flux of the star due to the large sample time, the
detectability $\Xi$ is close to 0. However, the detectability could
possibly be dramatically increased by moving away from opposition such
that $v_\mathrm{T}\Delta t \sim W^\ast$, allowing the event to be
detected. This is illustrated in Figure~\ref{fig:oppkbo}, which shows
lightcurves for a $r = 0.5$~km KBO at 40~AU measured at opposition and
at an opposition angle of $75^\circ$. At opposition, the detectability
of the event is $\Xi = 1.2 \times 10^{-4}$, but moving to $\phi =
75^\circ$ increases the detectability to $\Xi = 5.1 \times 10^{-3}$,
indicating that such an event is much more likely to be detected away
from opposition.

\section{Detection Sensitivity Limits for Photometric Systems}
\label{sec:threshold}
The detectability parameter described in the previous section can be
used in conjunction with the signal-to-noise ratio of a given survey
to define the sensitivity of the survey to objects of various sizes and
distances.  Consider the measured lightcurve consisting of a set of
measured photon counts at a detector $\tilde{N}_{i}=n\Delta t
\tilde{I}(t_{i})$, where $n$ is the photon rate of the background
star.  (Variables with a tilde indicate that they are measured and
contain an error term, for instance, $\tilde{I}(t_{i}) = I(t_{i}) +
e_{i}$ where $e_{i}$ is a random variable with zero mean, and variance
$\sigma^2$.)

If the null hypothesis of a constant lightcurve in which no
occultation event is present is selected, then the chi-squared
value over a set of $m$~points is
\begin{equation}
\chi^2 = \sum_{i=1}^{m}\frac{1}{\sigma^2}
\left[\tilde{N}_{i} - n\Delta t\right]^2.
\end{equation}

The sensitivity of a given survey is determined by calculating the
expected chi-squared value $\langle\chi^2\rangle$ and its
corresponding probability $P(\chi^2 > \langle\chi^2\rangle)$, and
asserting that detection of an event requires that the null-hypothesis
is rejected at a confidence level $p$.  To do so $\tilde{N}_{i}$ is
replaced with its definition given above and the expectation value is
taken.  The expected value of $\chi^2$ for an observed signal then
becomes
\begin{equation}
\langle\chi^2\rangle = \frac{n^2 \Delta t^2}{\sigma^2}\sum_{i=1}^{m}
\left[\left(\tilde{I}(t_{i}) - 1\right)^2\right] + m,
\end{equation}
where $\tilde{I}(t_{i})$ is the observed normalized lightcurve.
This can be expressed in terms of $\Xi$ using Equation~\ref{eq:detect_sum}:
\begin{equation}\label{eq:chi2_expected}
\langle\chi^2\rangle = \frac{n^2 \Delta t}{\sigma^2}\Xi + m.
\end{equation}
Note that the expectation value of this calculation with no event is
given by
\begin{equation}
\langle\chi^2\rangle=m,
\end{equation}
where $m$, the number of degrees of freedom, represents the underlying
variance due to noise. Therefore, we claim an occultation event is
\emph{detectable} if $\langle\chi^2\rangle$ is sufficiently larger
than $m$, such that $P(\chi^2 > \langle\chi^2\rangle) < p$.

For pure Poisson statistics, $\sigma^2=n\Delta t$ and therefore,
Equation~\ref{eq:chi2_expected} simplifies to
\begin{equation}
\langle\chi^2\rangle = n\Xi + m.
\end{equation}
This expression represents the best case scenario for any survey in
detecting an occultation curve.

Carefully considering the number of points $m$ over which $\chi^2$ is
calculated will help maximize $\langle\chi^2\rangle$ relative to $m$.
Note that from the definition of $\Xi$ in
Equation~\ref{eq:detect_sum}, the summand does not significantly
contribute to the sum beyond the occultation event duration
(Equation~\ref{eq:duration}) where $I(t_{i}) \approx 1$.  Therefore
the number of points needed to sufficiently sample an occultation
event should span the occultation event such that
\begin{equation}
m \approx \frac{W^{\ast}}{v_{T}\Delta t} .
\end{equation}  

The optimal value for $p$ depends on the survey statistics of a given
star and the target population, but it should be large enough to allow
sensitivity to as many events as possible but small enough to minimize
the false positive rate. A reasonable value should roughly be of order
$1/M$ where $M$ is the total number of observations of a background
source. A reasonable value of the threshold for the TAOS survey is $p
= 10^{-11}$ \citep{Lehner2006}, and we will adopt this value for the examples that
follow.

\begin{figure}
  \epsscale{1.0} \plotone{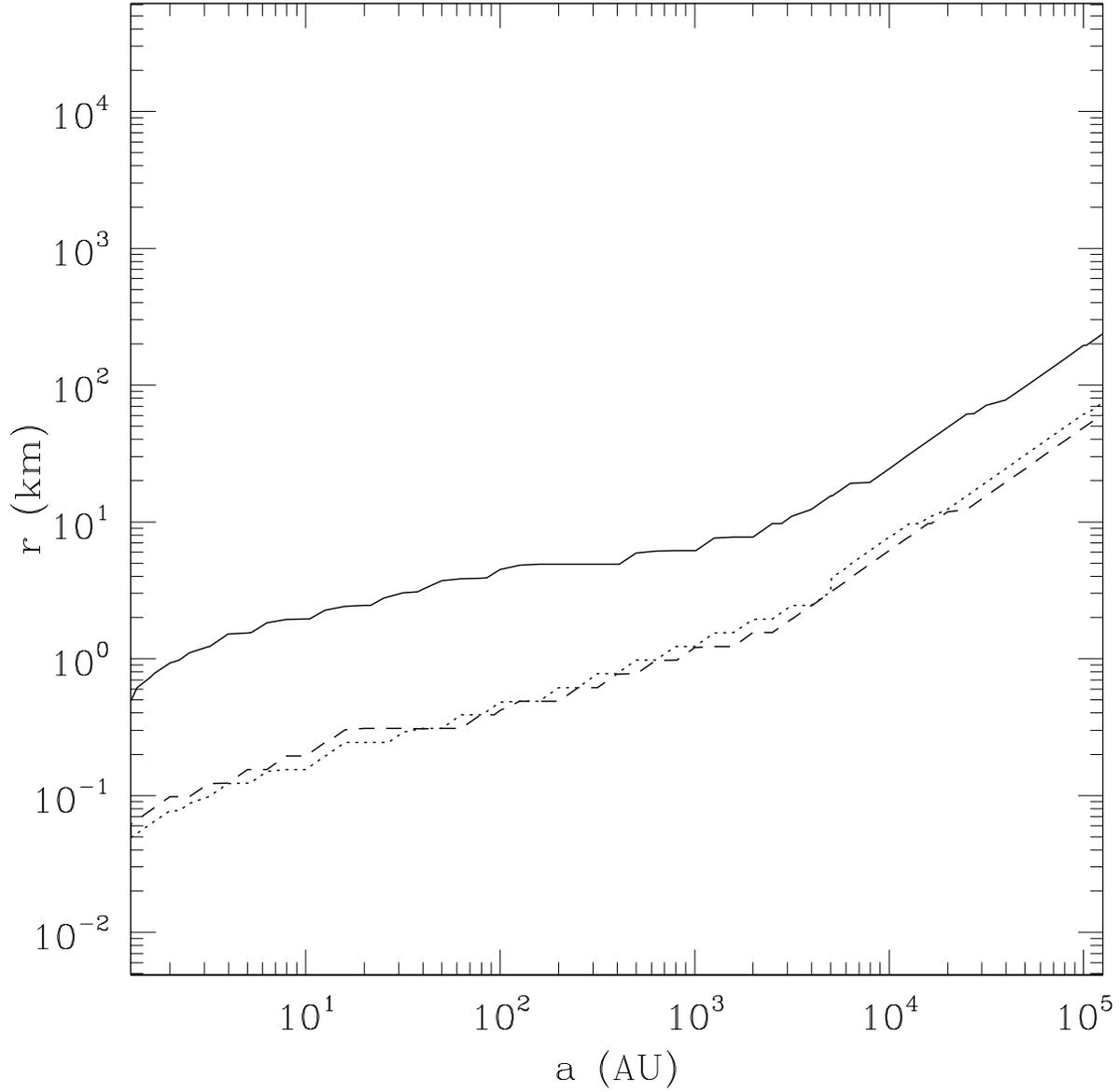}
  \caption{A plot of $P(\chi^2 > \langle\chi^2\rangle) = 10^{-11}$ for
    each of three telescope systems indicating the lower limit
    sensitivity to occultation events in the $a$-$r$ plane.  TAOS is
    indicated by the solid-line.  The dashed and dotted-lines
    respectively indicate the limits for the MMT and Whipple systems.}
  \label{fig:detect_lim} 
\end{figure}

Equation~\ref{eq:chi2_expected} can be written in terms of the
signal-to-noise ratio $S$ by noting that for a constant
source, $S=n\Delta t/ \sigma$.  Given this, and the above
approximation for $m$, we can compute $\langle\chi^2\rangle$ for a
given photometric system:
\begin{equation}\label{eq:chi2_expected_sn}
\langle\chi^2\rangle = \frac{S^2}{\Delta t}\Xi +
\frac{W^{\ast}}{v_{T}\Delta t}.
\end{equation}
Sensitivity limits for the three systems (TAOS, Whipple, and the MMT)
discussed in \S~\ref{sec:noise} are plotted in
Figure~\ref{fig:detect_lim}, given $p=10^{-11}$. This plot indicates
that TAOS is sensitive to Fresnel occultation events in the Kuiper
Belt down to objects of $r \sim 2$~km, as well as geometric events in
the more distant Extended Disk and Oort Cloud.  Looking at the
photometric sensitivity limits plotted for the MMT and Whipple
systems, we see very little difference between a ground-based survey
on a high signal-to-noise telescope like the MMT versus Whipple.
However, photometric performance is only one requirement of a
survey. The proposed Whipple system would follow more than 140,000
stars simultaneously, compared to less than 400 at the MMT. The
dedicated, 24-hour usage of Whipple would result in a vastly greater
number of detections.

\begin{figure}
  \epsscale{1.0} \plotone{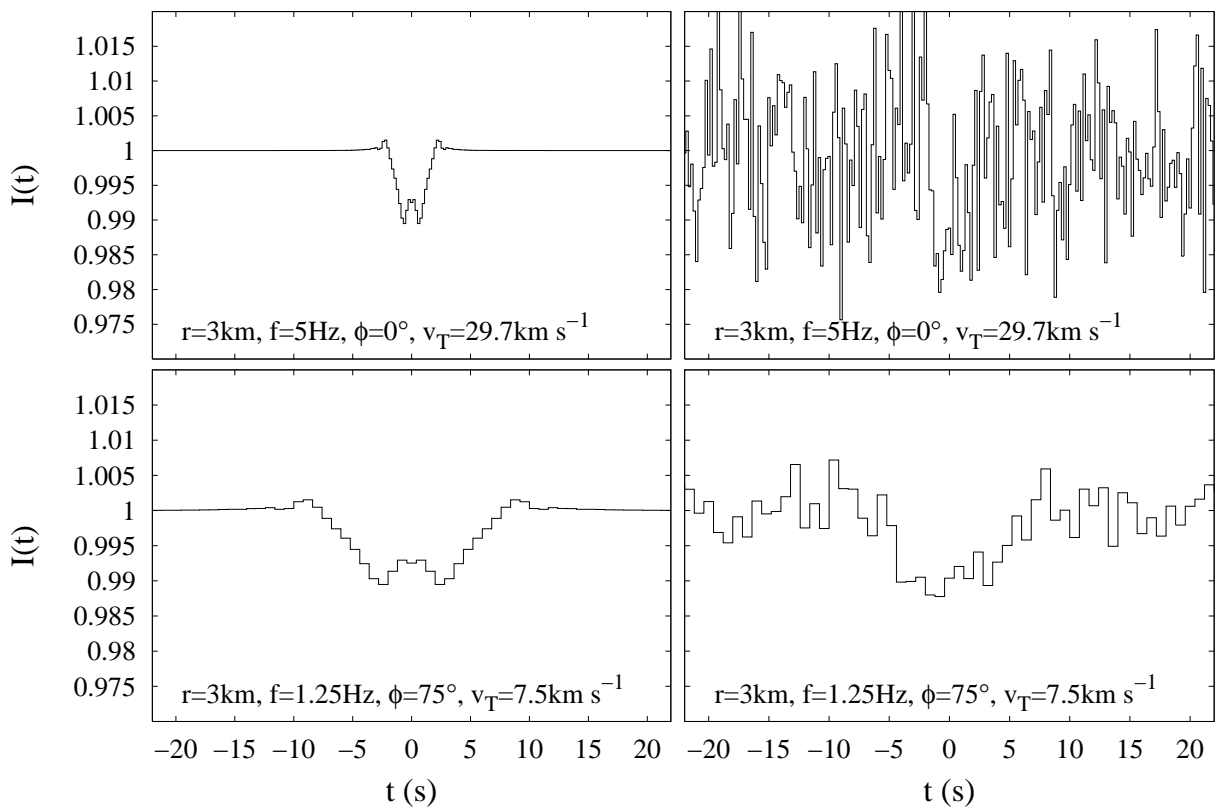}
  \caption{Lightcurves for a $r = 3$~km Oort Cloud object at
10,000~AU measured at opposition and at an opposition angle of
$75^\circ$. The lightcurve at opposition is sampled at 5~Hz, while the
lightcurve at $\phi = 75^\circ$ is sampled at 1.25~Hz. The panels on
the left show the theoretical lightcurve, while the lightcurves on the
right show lightcurves with simulated noise added for the proposed
Whipple survey.}
  \label{fig:oppoco} 
\end{figure}

As discussed in \S~\ref{sec:detect}, moving away from opposition can
significantly increase the detectability of an event if $W \ll
v_\mathrm{T}\Delta t$. However, even if $W \gg v_\mathrm{T}\Delta t$,
the detectability increases simply due to the larger event duration
$T$. In this case, the signal-to-noise ratio $S$ can be increased as
well by increasing the sample time accordingly. This is illustrated in
Figure~\ref{fig:oppoco}, which shows simulated lightcurves for the
Whipple survey for a $r = 3$~km Oort Cloud object at 10,000~AU
measured at opposition and at an opposition angle of $75^\circ$. The
relative velocity at $\phi = 75^\circ$ is a factor of four lower than
at $\phi = 0^\circ$, and therefore both $T$ and $\Xi$ are increased by
a factor of four. At $\phi = 75^\circ$, we have reduced the sampling
frequency by a factor of four, which increases the signal-to-noise
ratio $S$ by a factor of about 3.2 while holding $m$ constant. We can
then calculate $\langle\chi^2\rangle = 40.5$ at opposition and
$\langle\chi^2\rangle = 125.3$ at $\phi = 75^\circ$. Given the above
considerations, it is clear that the first term in
Equation~\ref{eq:chi2_expected_sn} increases by a factor of $3.2^2$
(both $\Xi$ and $\Delta t$ increase by a factor of four, so those
factors cancel), while the second term (the degrees of freedom)
remains constant ($v_\mathrm{T} \Delta t$ does not change). Thus the
value of $\langle\chi^2\rangle$ increases relative to the number of
degrees of freedom, making it more likely that the event can be
detected. In the example shown in Figure~\ref{fig:oppoco}, the
measured values of $\chi^2$ are 54.36 and 124.58 for $\phi = 0^\circ$
and $\phi = 75^\circ$ respectively, and with $m = 31$~degrees of
freedom, these values correspond to $P = 0.0058$ and $P = 3.7 \times
10^{-13}$. With a threshold value of $p = 10^{-11}$, the object could
be detected at $\phi = 75^\circ$, but not at opposition.

\section{Discussion}\label{sec:discuss}

In this paper, we have simulated occultation events in the Kuiper Belt
out to the Oort Cloud for three photometric systems incorporating the
effects of finite bandwidths and stellar spectra, finite source sizes,
and sampling.  Via these simulated occultation events, we have
quantitatively parameterized observed occultation events with the use
of the Fresnel scale, width $W^{\ast}$, depth $D$, and the
detectability $\Xi$, and we have shown how the detectability $\Xi$ can
be used to calculate the sensitivity of a survey using the
$\langle\chi^2\rangle$ parameter for an event.

In \S~\ref{sec:finite} lightcurve smoothing and occultation width
broadening due to finite source size effects were shown to be
significant factors in the detection of an event.  In
Figures~\ref{fig:detect_pts}~\&~\ref{fig:detect_fns} it was shown that
in spite of the augmentation of the occultation width by a finite
source radius, dampening of occultation event variation due to
background source smoothing lowers the detectability of events at
larger distances beyond the point of intersection between the
background source radius and the Fresnel scale.  Therefore monitored
background stars significantly minimize the detectability of an
occultation event when $r_{\ast} > F$.  Selection of background target
stars should aim to minimize the stellar size relative to the Fresnel
scale while balancing the need for decent signal-to-noise.  This
becomes a greater challenge at larger distances $a$ because of the
linear dependence of the projected stellar radius $r_{\ast}$ on $a$.
Even a relatively blue star of moderate brightness like an A0V V=12
star will lower the detectability at larger distances in the Outer
Solar System.  For the distant regions of the Solar System, the
ability to detect smaller objects will rely upon a photometry system's
limiting magnitude being relatively high.

In \S~\ref{sec:detect} and \S~\ref{sec:threshold}, a method of
determining a survey's sensitivity to objects in the Outer Solar
System regions by comparing the variation due to occultation events
$\Xi$ and the variation due to instrumental and photon count
fluctuations was presented.  In Figure~\ref{fig:detect_lim} it was
shown that even a ground survey of modest signal-to-noise is capable
of viewing Inner Oort Cloud objects as small as $r \sim 10$~km and
perhaps even Outer Oort Cloud objects with $r \sim 100$~km using the
occultation method.

The benefits of using larger telescopes with higher signal-to-noise
and the ability to sample at higher rates such as the described MMT
and Whipple campaigns, are evident.  The increase in detectability
gained in such cases would allow the surveys to push the lower limit of
detection for small objects in the Kuiper Belt even further almost by
an order of magnitude to objects with radii that are a few hundreds 
of meters.  Similarly the detection of smaller objects further
out in the Solar System is improved.

What can be detected by a survey in order to further aid the selection
of survey parameters can be further explored using our described
method.  Further work on incorporating the statistics of the
occultation event rates to determine the amount of necessary
observation time to significantly determine the population size
distribution of a set of Outer Solar System objects would contribute
greatly to the design and implementation of occultation surveys.

\acknowledgements

The efforts at Harvard University are supported by NASA grant
NNG05GA28G and NASA grant NNG05GO66G.

\end{document}